 \documentclass[authoryear,preprint,review,12pt]{elsarticle}

\usepackage{amssymb}
\usepackage{amsmath}
\usepackage{setspace}
\usepackage[authoryear,round]{natbib}
\usepackage{subcaption}
\usepackage{url}
\usepackage{float}
\usepackage{booktabs} 
\journal{Ocean Engineering}

\begin{document}

\begin{frontmatter}



\title{A Gradient Boosted Mixed-Model Machine Learning Framework for Vessel Speed in the U.S. Arctic}


\author[1]{Mauli Pant\corref{cor1}}
\ead{pantm2@vcu.edu}
\cortext[cor1]{Corresponding author.}

\affiliation[1]{organization={Integrative Life Sciences Doctoral Program, Center for Integrative Life Sciences Education, Virginia Commonwealth University},
            city={Richmond},
            state={Virginia},
            country={USA}}
\author[2]{Linda Fernandez}
\affiliation[2]{organization={School of Life Sciences and Sustainability, Department of Economics, Virginia Commonwealth University},
            city={Richmond},
            state={Virginia},
            country={USA}}

\author[3]{Indranil Sahoo}
\affiliation[3]{organization={Department of Statistical Sciences and Operations Research, Virginia Commonwealth University},
            city={Richmond},
            state={Virginia},
            country={USA}}


%% Abstract
\begin{abstract}
Understanding how environmental and operational conditions influence vessel speed is crucial for characterizing navigational conditions in the Arctic. We analyzed AIS data from 2010–2019 of vessel speed over ground (SOG). Over half the AIS records showed zero SOG, so treating zero and positive SOG as a continuous process can obscure important patterns. We therefore applied a two-stage machine learning framework, first modeling the probability of SOG > 0, and then modeling SOG conditional on being positive. We integrated AIS observations with sea ice concentration, course over ground, wind, bathymetric depth, distance to coast, vessel group, and navigational status. Gradient boosted decision trees with random effects captured nonlinear environmental responses and repeated observations. The positive SOG classifier achieved strong discrimination (AUC $\approx 0.85$), while the conditional speed model explained $\approx 77\%$ of out-of-fold variance. SHAP values quantified covariate effects by decomposing model predictions into additive contributions from individual variables. Distance to coast and bathymetric depth were dominant determinants of the likelihood and magnitude of vessel speed, while effect of changes in course, vessel group, and navigational status was secondary. Wind and sea ice effects were modest. These results empirically characterize Arctic vessel operating regimes for speed management and corridor level assessment.
\end{abstract}


% \begin{highlights}
% \item Environmental and operational drivers of zero and positive vessel speeds in Arctic
% \item SHapley Additive exPlanation enables interpretable assessment of large scale AIS data
% \item Results show impact of distance to coast, depth, and change in COG on vessel speed
% \end{highlights}

%% Keywords
\begin{keyword}
Arctic shipping; Automatic Identification System (AIS); Environmental drivers; Gaussian Process; GPBoost; Spatial effects; Speed Over Ground (SOG).

\end{keyword}

\end{frontmatter}



\section{Introduction}\label{Intro}

The Arctic presents one of the most challenging maritime environments in the world.
Shifting sea ice, high winds, and complex bathymetric depth create conditions where safe and
efficient navigation requires constant attention. These challenges are intensifying as
climate change reduces sea ice extent, increasing seasonal vessel traffic while raising
concerns about navigational safety. According to the ten-year forecast report \citep{CMTS2019}, Arctic
vessel traffic could be three times higher in 2030 than in 2008. Because vessel speed influences navigation risk,
exposure time in sensitive habitats, and the severity of environmental effects,
understanding the drivers of observed speed over ground (SOG) is central to Arctic
navigation management. 

Much of the existing literature emphasizes analytical modeling approaches, with
comparatively few studies leveraging empirical or statistical analyses of observed vessel
navigation \citep{Aksenov2017Navigability, Chang2015-hr, Goerlandt2017-kc, Kuuliala2017-mz,
Loptien2014-mu, Montewka2015-xz, Pruyn2016-xt}. In addition, environmental drivers are typically treated in isolation rather than jointly. This limits the understanding of
combined effects of sea ice, wind, and bathymetric depth on vessel speed over ground (SOG), an operational variable central to fuel efficiency, navigational safety, and environmental impact.

Our contribution to the literature involved analyzing Automatic Identification System (AIS) vessel observations from 2010–2019 during the open-water season (July–October),
integrated with satellite derived environmental data. By explicitly separating zero and positive speed regimes and modeling their environmental and operational drivers, this study provided an interpretable characterization of Arctic vessel operations and identified where speed related navigational and environmental risks were most likely to arise. The results offer a data driven basis for understanding how spatial context and steering shape vessel speed under increasing Arctic vessel traffic.

While ice-related speed thresholds provide a useful baseline for identifying potentially hazardous operating conditions (Figure~\ref{fig:Access}), observed vessel speeds in the
U.S. Arctic exhibited substantial variability even under similar ice conditions (Figure~\ref{fig:Wholeplt}). This variability indicates that vessel speed decisions during
the open water navigation season were shaped by a broader set of environmental and operational constraints beyond ice alone. In particular, bathymetric depth, proximity to
shore, wind conditions, and vessel specific navigational status interacted to influence SOG across space and time. As a result, threshold based ice risk metrics, while operationally
informative, cannot capture the nonlinear and multivariate drivers of Arctic vessel navigation observed in practice.

This study addressed these limitations by embedding ice related navigation risk within a
data driven modeling framework that explicitly accounted for environmental, spatial, and
operational drivers of vessel speed. To this end, we employed a tree based gradient boosting
framework with structured random effects. This modeling approach is well suited for
capturing nonlinear relationships, threshold effects, and heterogeneous vessel operations
in large scale AIS data \citep{chen2016xgboost,ke2017lightgbm,Sigrist2021}. By integrating
more than a decade of AIS observations with environmental covariates, our approach extended
existing ice risk frameworks from static metrics toward a statistical characterization of
vessel navigation under varying environmental conditions.

Tree-based machine learning models have been widely applied to AIS data for vessel
navigation analysis, including vessel speed prediction, characterization of speed
distributions, and identification of low speed or zero speed operating contexts in coastal
waters \citep{Ma2024AISCoastalSpeed,WANG2022110691,YANG2024103426}. These studies demonstrate
the predictive flexibility of ensemble tree methods but typically treat vessel speed as a
single continuous process and do not explicitly address the strong zero inflation in SOG.
Such approaches also rarely account for repeated observations within vessels, locations, or
time periods, limiting their utility for policy relevant interpretation of vessel operating
regimes. Furthermore, environmental drivers such as sea ice, bathymetric depth, and wind are
often incorporated only in limited form or excluded entirely, constraining physical
interpretation of model outputs.

This study advances the existing literature by introducing a unified framework that treats zero and positive vessel speeds as distinct operating regimes rather than a single continuous process. By separating stationary behavior (SOG = 0) from active navigation (SOG > 0), the analysis distinguishes stopping, maneuvering, and sustained transit patterns that are otherwise conflated in AIS-based speed models. Environmental and spatial drivers, including sea ice concentration, bathymetric depth, distance to coast, and wind forcing, are incorporated jointly, allowing nonlinear responses and interactions to be resolved across operating regimes. The framework further accounts for repeated observations within vessels, spatial grid cells, and daily time periods through structured random effects, capturing persistent operational and geospatial constraints on vessel speed.

To interpret the fitted machine learning models, we employ SHapley Additive exPlanations \citep[SHAP;][]{Lundberg2017SHAP} framework, which takes a game theoretic approach to explaining the output of a machine learning model. Shapley values provide a principled framework for fairly attributing a model’s prediction to individual features by averaging their marginal contributions across all possible feature coalitions. When applied to machine learning models, this yields additive, locally consistent explanations that decompose each prediction relative to a baseline, enabling interpretable assessment of nonlinear effects and feature interactions in complex black-box models \citep{wikle2023illustration}. This approach facilitates examination of environmental and operational drivers of vessel speed across heterogeneous Arctic navigation contexts using large-scale AIS data, characterizing observed vessel speeds as the outcome of interacting environmental, spatial, and operational processes.

\section{AIS Vessel Traffic Data and Environmental Covariates}\label{sec:data}

Vessel traffic data are transmitted via an onboard Automatic Identification System (AIS)
transponder, which communicates vessel location and identity to other vessels and coastal
authorities. These data are recorded for vessels navigating U.S. coastal waters and are
distributed by the U.S. Coast Guard through the AIS platform
(\url{https://marinecadastre.gov/AIS/}), which is jointly managed by the Bureau of Ocean
Energy Management (BOEM), the National Oceanic and Atmospheric Administration (NOAA), and
the U.S. Coast Guard Navigation Center. AIS records are available from 2009 onward; in this
study, we focused on the period 2010--2019.

Analysis was restricted to July--October, coinciding with the Arctic shipping season, and
to a spatial extent defined by longitude $(-138^\circ, -172^\circ)$ and latitude
$(66.6^\circ, 72.5^\circ)$ within the U.S. Arctic.  Missing
values in vessel attributes were supplemented where possible; residual gaps were imputed
using vessel type means. Across 2010--2019, a total of 14{,}789{,}591 AIS observations were
retained for July--October within the defined spatial extent. Tug/tow vessels accounted
for the highest volume of observations, followed by cargo and other categories. Cruise
ships exhibited the highest average SOG, whereas cargo, tanker, and search and rescue
(SAR) vessels represented the largest and heaviest groups.

The empirical distribution of SOG was highly left skewed, with the majority of observations
concentrated near zero. More than half (53\%) of all AIS records reported a speed of exactly
0~knots, indicating strong zero inflation in vessel SOG. This characteristic has direct
implications for model specification, as the data combine frequent zero SOG observations
with a continuous distribution of positive SOG values. Table~\ref{tab:vesselgroup_movement}
summarizes the composition of AIS observations by vessel group conditional on zero or
positive SOG. Tug and tow vessels dominated both categories, reflecting their extensive
presence in the study region, but accounted for a substantially larger share of zero SOG
observations. In contrast, cargo vessels represented a higher proportion of positive SOG
observations. High speed craft were disproportionately represented among zero SOG
observations.

Additional patterns emerged when AIS records were stratified by navigation status
(Table~\ref{tab:status_movement}). Most positive SOG observations corresponded to vessels
under way using engine, while zero SOG observations were more frequently associated with
anchoring, mooring, and undefined operational statuses. Together,
Tables~\ref{tab:vesselgroup_movement} and \ref{tab:status_movement} show that zero SOG AIS
observations spanned many vessel groups and navigation statuses. This heterogeneity has
important implications for modeling vessel speed and motivated the use of a two-stage
framework that distinguishes zero SOG observations from positive SOG values.

\begin{figure}[H]
  \centering
  \includegraphics[width=.8\textwidth]{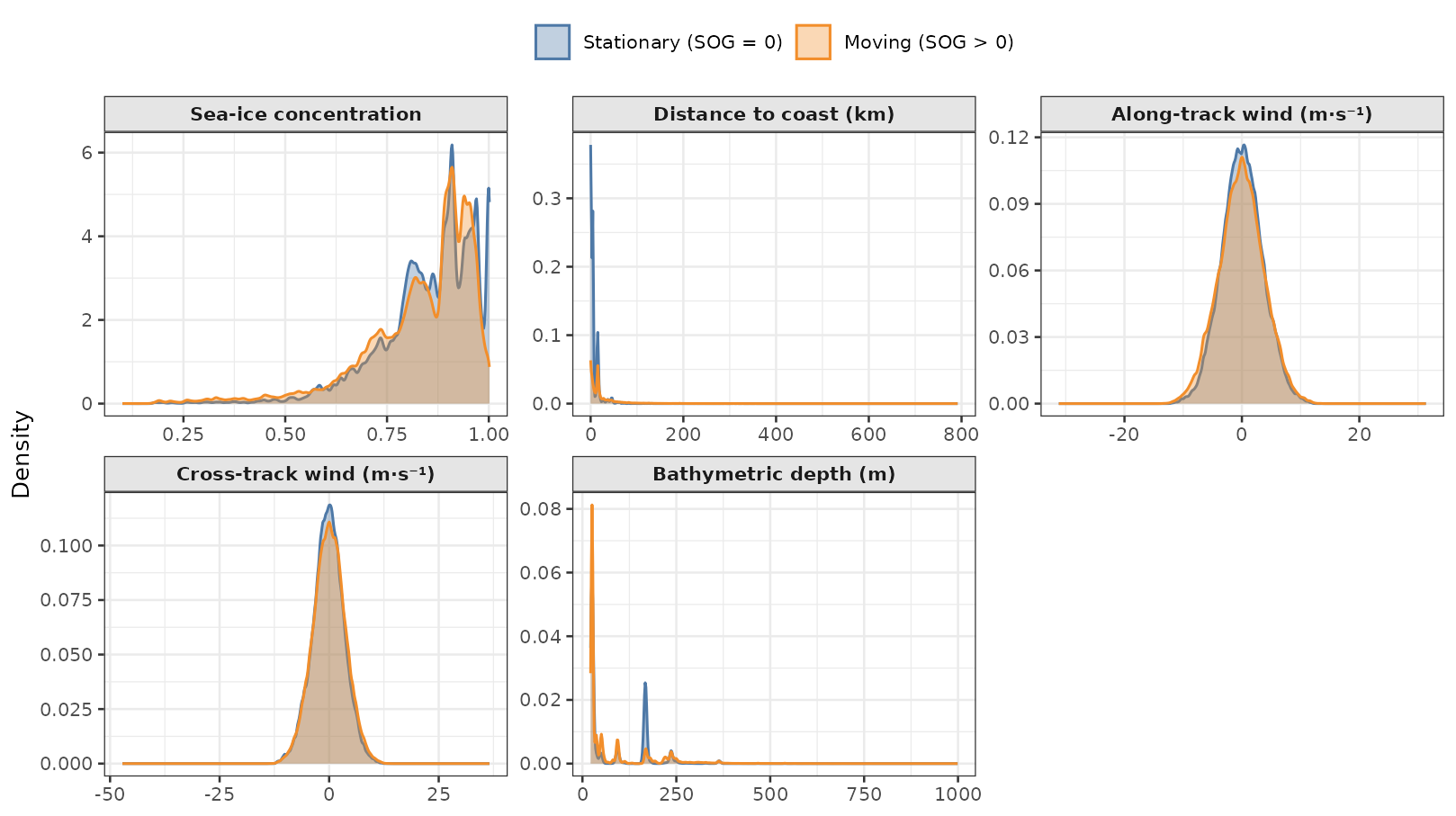}
  \caption{Distribution of environmental and spatial covariates by Zero or positive SOG (July-October, 2010-2019). Density curves compare SOG = 0 and SOG $>$ 0 AIS observations. Wind effect is represented using along-track and cross-track components (m$\cdot$s$^{-1}$), consistent with the regression specification. Axes are shown on variable specific scales to emphasize within variable distributional differences.}
  \label{fig:covariate_density_movement}
\end{figure}

Figure~\ref{fig:covariate_density_movement} shows that sea ice concentration was high for
both zero SOG and positive SOG observations. In contrast, distance to coast clearly
differentiated the two, with zero SOG observations concentrated much closer to shore. Wind
conditions were similar for zero SOG and positive SOG observations, although their
distributions differed modestly. Vessels with both zero and positive SOG were observed
across shallow and deep waters.
\begin{table}[h]
\centering
\caption{Distribution of AIS observations by vessel group and Zero Inflation (July-October, 2010-2019). Percentages are calculated within each category ($SOG=0$ or $SOG>0$).}
\label{tab:vesselgroup_movement}

\begin{tabular}{lrrrr}
\toprule
\textbf{Vessel Group} &
\multicolumn{2}{c}{\textbf{Positive (SOG $>$ 0)}} &
\multicolumn{2}{c}{\textbf{Zero (SOG = 0)}} \\
\cmidrule(lr){2-3}\cmidrule(lr){4-5}
 & \textbf{Count} & \textbf{\%} & \textbf{Count} & \textbf{\%} \\
\midrule
Tug / Tow        & 4,401,128 & 63.23 & 5,791,372 & 73.97 \\
Cargo            & 1,051,955 & 15.11 &   921,494 & 11.77 \\
Other            &   971,605 & 13.96 &   404,417 &  5.17 \\
Fishing          &   209,847 &  3.02 &   132,906 &  1.70 \\
High-Speed Craft &    38,486 &  0.55 &   346,506 &  4.43 \\
Tanker           &    78,541 &  1.13 &   108,931 &  1.39 \\
Pleasure Craft   &    68,084 &  0.98 &    12,603 &  0.16 \\
Passenger        &    19,872 &  0.29 &     9,424 &  0.12 \\
Cruise Ship      &     5,253 &  0.08 &    13,636 &  0.17 \\
SAR              &    12,712 &  0.18 &       328 &  0.00 \\
Unspecified      &    88,392 &  1.27 &    65,646 &  0.84 \\
Other Specialized$^\dagger$ & 15,526 & 0.22 & 22,280 & 0.29 \\
\bottomrule
\end{tabular}

\begin{flushleft}
{\footnotesize
$^\dagger$Includes anti pollution vessels, dredgers, pilot vessels, service ships, reserved categories, and other low frequency vessel types.
}
\end{flushleft}
\end{table}

\begin{table}[H]
\centering
\caption{Distribution of AIS observations by navigation status and Zero Inflation  (July-October, 2010-2019). Percentages are calculated within each category($SOG=0$ or $SOG>0$).}
\label{tab:status_movement}

\begin{tabular}{lrrrr}
\toprule
\textbf{Navigation Status} &
\multicolumn{2}{c}{\textbf{Positive (SOG $>$ 0)}} &
\multicolumn{2}{c}{\textbf{Zero (SOG = 0)}} \\
\cmidrule(lr){2-3}\cmidrule(lr){4-5}
 & \textbf{Count} & \textbf{\%} & \textbf{Count} & \textbf{\%} \\
\midrule
Status 0  & 5,520,374 & 79.32 & 5,949,981 & 75.99 \\
Status 1  &   488,561 &  7.02 &   482,709 &  6.17 \\
Status 3  &   374,897 &  5.39 &   198,430 &  2.53 \\
Status 15 &   191,891 &  2.76 &   552,046 &  7.05 \\
Status 5  &   146,743 &  2.11 &   315,647 &  4.03 \\
Status 11 &   124,458 &  1.79 &   228,830 &  2.92 \\
Status 12 &    56,595 &  0.81 &    49,357 &  0.63 \\
Status 8  &    49,176 &  0.71 &    27,012 &  0.35 \\
Status 2  &     3,310 &  0.05 &    21,058 &  0.27 \\
Status 6  &     2,272 &  0.03 &     4,459 &  0.06 \\
Status 7  &     1,239 &  0.02 &        -- &   --  \\
Status 10 &       532 &  0.01 &        14 &  0.00 \\
\bottomrule
\end{tabular}

\begin{flushleft}
{\footnotesize
AIS navigation status codes follow the standard AIS specification: Status~0 = under way using engine; Status~1 = at anchor; Status~2 = not under command; Status~3 = restricted maneuverability; Status~4 = constrained by her draught; Status~5 = moored; Status~6 = aground; Status~7 = engaged in fishing; Status~8 = under way sailing; Status~9 = reserved for future amendment; Status~10 = power-driven vessel towing astern; Status~11 = power-driven vessel towing alongside; Status~12 = reserved for future use; Status~13 = reserved for future use; Status~14 = AIS-SART (search and rescue transmitter); Status~15 = not defined. Rare navigation status categories ($<0.05\%$) are retained for completeness.
}
\end{flushleft}
\end{table}

The following environmental variables were extracted and collocated with AIS positions on a
$0.5^\circ \times 0.5^\circ$ grid at daily resolution:
\begin{itemize}
\item \textbf{Sea ice concentration.}
Daily sea ice concentration fields were obtained from the NOAA OISST/DOISST high-resolution blended analysis NetCDF products (icec.day.mean.YYYY.nc; variable icec) for 2010–2019. These fields are distributed with the OISST/DOISST product and use external sea-ice analyses (NCEP and NASA GSFC), as documented in the dataset metadata and described by \citep{Huang2021DOISST}.
Ice concentration was reported on a latitude--longitude grid at a  spatial
resolution of 0.25$^\circ$ and expressed as a unitless fractional ice cover (0--1).
NetCDF time stamps were converted to calendar dates using the native time encoding
(days since 1800-01-01, UTC), and longitudes were wrapped to $[-180^\circ,180^\circ]$
for consistency with AIS coordinates.
Ice fields were subset to the U.S. Arctic study domain
(66.6--72.5$^\circ$N; 172--138$^\circ$W) and subsequently aggregated to a 0.5$^\circ$
grid using spatial averaging to align with the resolution of environmental covariates
mapped to vessel observations.

Sea ice concentration was collocated with AIS positions by assigning each AIS record to the
nearest daily ice date and extracting ice values at vessel locations. For each day, the ice
field was rasterized to a $0.5^\circ \times 0.5^\circ$ grid (EPSG:4326) using the mean ice
concentration within each grid cell, and vessel level ice values were obtained by spatial
extraction from this raster. When extraction returned missing values (e.g., due to gaps in
the gridded ice field), missing ice values were imputed using a two step geostatistical
procedure. First, missing values in the daily ice field were filled using inverse distance
weighting (IDW; $n_{\max}=10$, power parameter $p=2$). Second, an empirical variogram was
estimated from the filled daily ice field and a spherical variogram model was fitted; ordinary
kriging with $n_{\max}=10$ neighbors was then used to predict ice concentration at AIS
locations for that day. This procedure ensured complete ice covariate coverage for all AIS
observations used in the modeling.

\item \textbf{Wind and bathymetry.}
Daily gridded fields of 10-m wind components and bathymetric depth \citep{era5_single_levels} were collocated
with AIS positions using the same day matching procedure applied to sea ice
concentration. Environmental fields were provided at daily temporal resolution and
mapped to a common $0.5^\circ \times 0.5^\circ$ latitude--longitude grid
(EPSG:4326), matching the spatial resolution used for AIS 
covariates. For each day, AIS time stamps were assigned to the nearest available
daily environmental field.

Wind direction was represented by the eastward and northward 10-m wind components,
denoted $u_{10}$ and $v_{10}$ (m\,s$^{-1}$), and bathymetric depth was denoted
$\mathrm{wmb}$ (m). For each day, environmental fields were rasterized to the
$0.5^\circ \times 0.5^\circ$ grid using mean aggregation within grid cells, and
initial vessel level covariate values were obtained by bilinear extraction from
the daily rasters.

To ensure complete covariate coverage at vessel locations and to reduce the effects of assigning identical grid-cell values in the daily environmental fields, wind and bathymetry were interpolated to vessel positions using ordinary kriging,applied
separately for $u_{10}$, $v_{10}$, and $\mathrm{wmb}$ on each day. For each variable,
duplicated spatial points were removed, an empirical variogram was computed, and a
parametric variogram model was fitted prior to kriging to AIS locations using up to
$n_{\max}=10$ neighboring points. An exponential variogram model was used for
$u_{10}$, while Gaussian variogram models were used for $v_{10}$ and
$\mathrm{wmb}$. The resulting kriged predictions were assigned as the final
vessel level covariate values for that day.
\end{itemize}

Interpolation was used solely to ensure complete covariate coverage and did not introduce
new vessel observations. 

\begin{figure}[H]
    \centering
    \includegraphics[width=.8\textwidth]{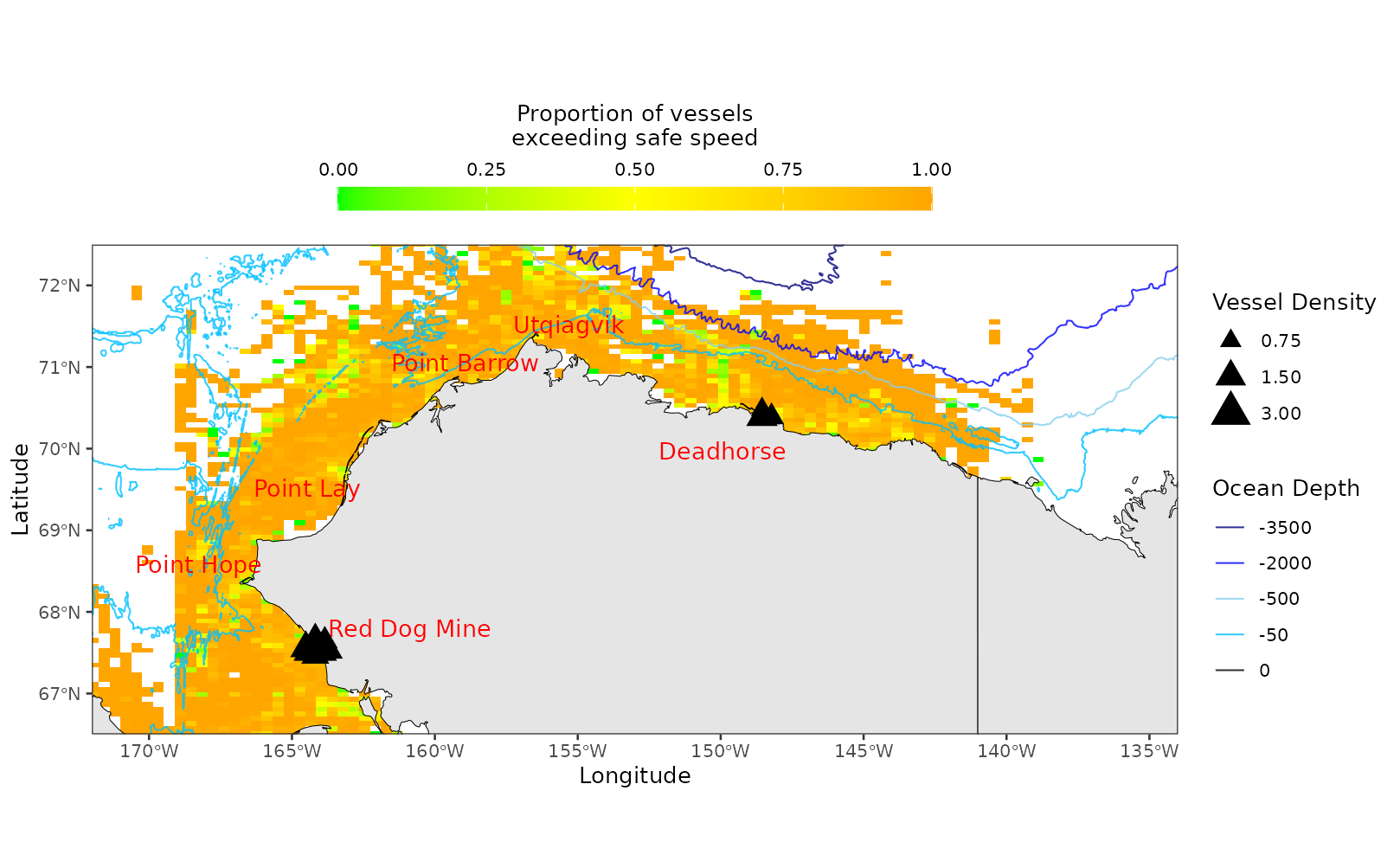}
    \caption{Spatial distribution of binary ice-related speed risk across the U.S. Arctic (July–October, 2010–2019). Grid cell shading indicates the proportion of AIS observations exceeding an ice-dependent safe speed threshold (Figure A1). Black triangles mark the ten highest vessel density locations, scaled by a relative density index. Bathymetric depth contours provide spatial context.
}
    \label{fig:Wholeplt}
\end{figure}

Figure~\ref{fig:Wholeplt} presents a synthesized snapshot of the integrated datasets, illustrating areas of heightened navigational concern where all three datasets overlap within the US Arctic. This visualization highlights the integrative spatial framework developed for this study, which combines vessel volume, environmental conditions, and ecological factors in a unified analysis.

Spatial overlays of vessel observations, sea ice concentration, and wind conditions show
substantial co-occurrence in coastal corridors where persistent ice intersects with shallow
shelves and steep depth gradients (Figure~\ref{fig:Wholeplt}). We further calculated a
binary ice risk indicator by comparing each vessel’s observed SOG with a reference safe
speed curve derived from the ACCESS (Arctic Climate Change Economy and Society) report
\citep{HSVA2014FuelConsumption} (see Figure~\ref{fig:Access}), which relates recommended
operating speed to sea ice concentration. Risk was assigned a value of 1 when a vessel’s
SOG exceeded the ACCESS derived safe speed for the corresponding ice concentration, and 0
otherwise. This binary indicator was then aggregated within $0.5^\circ \times 0.5^\circ$
grid cells to estimate the proportion of potentially risky vessel transits at each
location. Spatial overlap between elevated ice risk, high vessel density, and shallow
bathymetric depth highlights areas where navigational and ecological risks may co-occur,
particularly in regions that overlap with sensitive marine mammal habitats
\citep{Reeves2014,Hauser2018,Halliday2020}.

\section{Statistical Methodology: A Two-Stage Machine Learning Framework}

 % Modeling SOG as a single continuous response ignores the distinct processes generating zero and positive speeds and can reduce interpretability. To address this issue, 
 Vessel SOG was modeled using a two-stage framework consisting of
(i) classification of whether a vessel's SOG was zero or positive and
(ii) conditional regression of vessel speed given positive SOG.
The two stages were estimated sequentially rather than under a single joint likelihood,
separating the stochastic process governing the presence or absence of speed from the
process governing speed variability during active navigation.
 % This formulation separates the processes governing whether a vessel is moving from those governing how fast it moves, thereby improving interpretability and statistical stability.

\subsection{GPBoost Modeling Framework} \label{sec:gpboost}

Both stages of the analysis were implemented using the \texttt{GPBoost} framework
\citep{Sigrist2021}, which combines gradient boosted decision trees (GBDTs) with grouped
Gaussian random effects. Random effects captured correlation from repeated observations
within the same vessel (MMSI), grid cell, and day (TimeID), rather than assuming
independence. Spatial grouping used a regular $0.5^\circ \times 0.5^\circ$
latitude--longitude grid to align with the resolution of environmental covariates and
ensure consistent spatial coverage where data were missing. Temporal grouping was defined
at the daily level (\texttt{TimeID}) to account for day-to-day correlation in vessel
navigation. This structure allowed separation of systematic covariate effects from
vessel specific, spatial, and day to day variability, reducing bias that would arise from
treating repeated observations as independent. Further, GPBoost provided a unified
modeling architecture for flexible nonlinear covariate effects while explicitly accounting
for correlated structure arising from repeated vessel observations across space and time.

For observation $i$, the latent linear predictor is written as
\[
\eta_i =
f(\mathbf{x}_i)
+ u_{g_{\mathrm{mmsi}}(i)}
+ s_{g_{\mathrm{cell}}(i)}
+ t_{g_{\mathrm{time}}(i)},
\]
where $f(\mathbf{x}_i)$ is an additive ensemble of regression trees, and
$u_{g_{\mathrm{mmsi}}(i)}$, $s_{g_{\mathrm{cell}}(i)}$, and $t_{g_{\mathrm{time}}(i)}$
are grouped Gaussian random effects for vessel, spatial grid cell, and time index,
respectively. The grouped random effects are modeled as
\[
u_v \sim \mathcal{N}(0, \sigma^2_{\mathrm{MMSI}}), \quad
s_c \sim \mathcal{N}(0, \sigma^2_{\mathrm{cell}}), \quad
t_\tau \sim \mathcal{N}(0, \sigma^2_{\mathrm{time}}),
\]
and the boosted tree and random effects components were estimated jointly using the
integrated likelihood approach implemented in \texttt{GPBoost}.
The nonlinear predictor is constructed as
\[
f(\mathbf{x}_i) = \sum_{m=1}^{M} \alpha\, T_m(\mathbf{x}_i),
\]
where $\alpha$ is the learning rate and each tree $T_m$ is fitted sequentially to residual
structure left by the previous ensemble.

The covariate vector is defined as
\[
\mathbf{x}_i =
\left(
\begin{aligned}
&\text{scaled\_icec}_i,\ \text{scaled\_wind\_along}_i,\ \text{scaled\_wind\_cross}_i,\\
&\text{scaled\_dist\_to\_coast}_i,\
\text{scaled\_bathy}_i,\
\text{scaled\_cog\_sin}_i,\\
&\text{scaled\_cog\_cos}_i,\ \text{scaled\_}\Delta\text{COG}_i,\
\text{VesselGroup}_i,\ \text{Status}_i
\end{aligned}
\right)^{\mathsf T}.
\]

Sea ice concentration, $\text{icec}_i$, denotes fractional ice cover at observation $i$
(unitless; 0--1), with $\text{scaled\_icec}_i$ its standardized form. Wind direction is
represented by the eastward and northward 10-m wind components ($u_{10}$, $v_{10}$), which
are rotated relative to each vessel’s course over ground (COG) to obtain along-track and
cross-track wind components (m\,s$^{-1}$). Distance to coast is measured in kilometers
(km), and bathymetric depth represents seafloor depth in meters (m).

Directional effects are represented using a circular encoding of course over ground via
its sine and cosine components, $\text{scaled\_cog\_sin}_i$ and
$\text{scaled\_cog\_cos}_i$, which preserve directional continuity at
0/360$^\circ$. Short-term steering intensity was captured by the absolute circular
change in course, $\text{scaled\_}\Delta\text{COG}_i$, measured in degrees between
successive AIS observations. Categorical variables $\text{VesselGroup}_i$ and
$\text{Status}_i$ encoded vessel type and navigational status, respectively.

All continuous variables were standardized to zero mean and unit variance, so covariates
are unitless and represent relative effects. Categorical variables were encoded as integer
indices, and boosted tree effect magnitudes therefore reflect relative importance rather
than original units.

% AIS datasets exhibit substantial zero inflation in SOG (SOG), with a large fraction of observations reporting
% exactly zero speed. These zero values do not arise from a single, well-defined operational state and may occur across a
% range of navigational statuses and contexts. However, zero-speed observations are statistically distinct from positive-speed
% observations in that they represent the absence of forward motion, whereas positive values reflect continuous transit speed.

\subsubsection{Step 1: Binary Classification of Zero and Positive SOG}
To account for structured dependence in AIS observations, random intercepts were specified
for 530 unique MMSI identifiers, 327 spatial grid cells, and 1{,}194 daily time indices. For
each AIS observation, SOG was used to define a binary indicator
\[
Y_i =
\begin{cases}
1, & \text{if } \mathrm{SOG}_i > 0, \\
0, & \text{if } \mathrm{SOG}_i = 0,
\end{cases}
\]
where $Y_i = 1$ denotes positive SOG and $Y_i = 0$ denotes zero SOG.

In this stage, GPBoost was fitted with a Bernoulli likelihood and a probit link,
\[
p_i = \Phi(\eta_i),
\]
where $\Phi(\cdot)$ denotes the standard normal cumulative distribution function and
$\eta_i$ is the latent linear predictor combining nonlinear covariate effects and grouped
random effects as defined in Section~\ref{sec:gpboost}. The probit link was used because it
is compatible with the latent Gaussian mixed-effects formulation underlying GPBoost
\citep{Sigrist2021}.

Across the full dataset, 7{,}829{,}481 AIS observations (52.9\%) corresponded to zero SOG
(SOG = 0), while 6{,}959{,}580 observations (47.1\%) indicated positive speeds (SOG $> 0$),
resulting in moderate class imbalance. Model performance was evaluated using stratified
$K$-fold cross-validation with random subsampling of the majority class to address class
imbalance. Let $\mathcal{D}=\{(x_i,y_i)\}_{i=1}^N$, with $y_i\in\{0,1\}$ indicating zero SOG
and positive SOG. Within each training fold $k$, a balanced training set was constructed as
\[
\tilde{\mathcal{D}}^{(k)}_{\text{train}}
=
\mathcal{D}^{(k)}_{\text{train},1}
\;\cup\;
\mathcal{S}\!\left(\mathcal{D}^{(k)}_{\text{train},0},
\;|\mathcal{D}^{(k)}_{\text{train},1}|\right),
\]
where $\mathcal{S}(\cdot)$ denotes random sampling without replacement. Validation folds
retained the original class proportions. This procedure follows standard practice in
imbalanced classification \citep{Japkowicz2002}.Majority class subsampling was applied within training folds to stabilize fitting given strong spatial clustering of zero SOG observations nearshore.

\subsubsection{Step 2: Conditional Regression of Vessel Speed}

The conditional SOG regression was fitted to the moving subset (SOG $> 0$), comprising
6{,}959{,}580 observations from 527 unique vessels (MMSI), 325 spatial grid cells, and
1{,}191 daily time indices (\texttt{TimeID}). These reduced counts reflect the exclusion of
zero-speed records, which removed vessels, grid cells, and days associated solely with zero
SOG observations.

Let $\mathrm{SOG}_i$ denote the observed SOG for AIS observation $i$. The continuous
response variable for the regression stage is defined as
\[
y_i = \sqrt{\mathrm{SOG}_i}, \qquad \text{for } Y_i = 1,
\]
where the square root transformation was applied to positive SOG values to mitigate right skewness and heteroskedasticity while maintaining a monotonic and physically interpretable relationship with vessel speed. This choice preserves relative differences among moderate and high speeds and avoids excessive compression associated with logarithmic transformations.
The transformed speed was modeled as
\[
y_i
=
f(\mathbf{x}_i)
+ u_{g_{\mathrm{mmsi}}(i)}
+ s_{g_{\mathrm{cell}}(i)}
+ t_{g_{\mathrm{time}}(i)}
+ \varepsilon_i,
\]
where $\varepsilon_i$ is an observation level error term with mean zero. After accounting
for covariate effects and vessel, space, and time  specific random effects, the remaining
observation level variation was assumed independent and reflected measurement noise and
other small scale, unmodeled influences. Tree depth, learning rate, and minimum leaf size
were selected using cross-validated performance on the training data, with conservative
regularization to prioritize generalization.

\section{Model Evaluation}

\subsection{Evaluation of Binary SOG Classification}

\begin{figure}[h]
  \centering  \includegraphics[width=.8\linewidth]{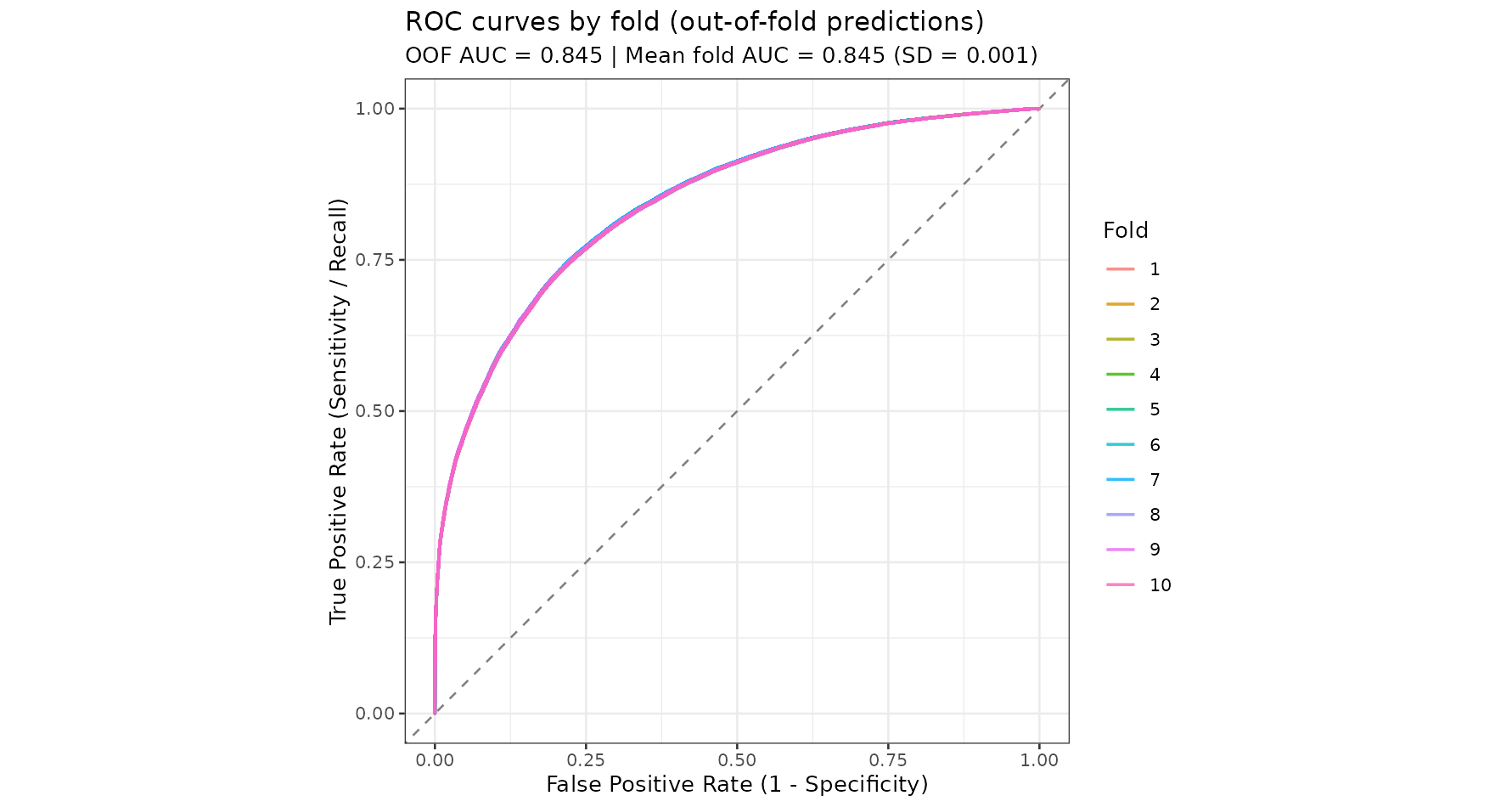}
  \caption{\normalsize ROC curve with AUC (out-of-fold predictions). The close overlap of curves across folds
indicates that classifier performance is robust to spatial, temporal, and
vessel level heterogeneity and generalizes well beyond individual training sets.}
  \label{fig:sog_roc}
\end{figure}

Model discrimination was assessed using the area under the receiver operating
characteristic curve (AUC) computed from out-of-fold predictions (Figure \ref{fig:sog_roc}). The overall
out-of-fold AUC was $0.845$, with fold level values ranging from $0.844$ to $0.846$
(standard deviation $9.2 \times 10^{-4}$), indicating strong and highly stable
discrimination between zero SOG and positive SOG observations.

\begin{figure}[h]
\centering
\includegraphics[width=.7\linewidth]{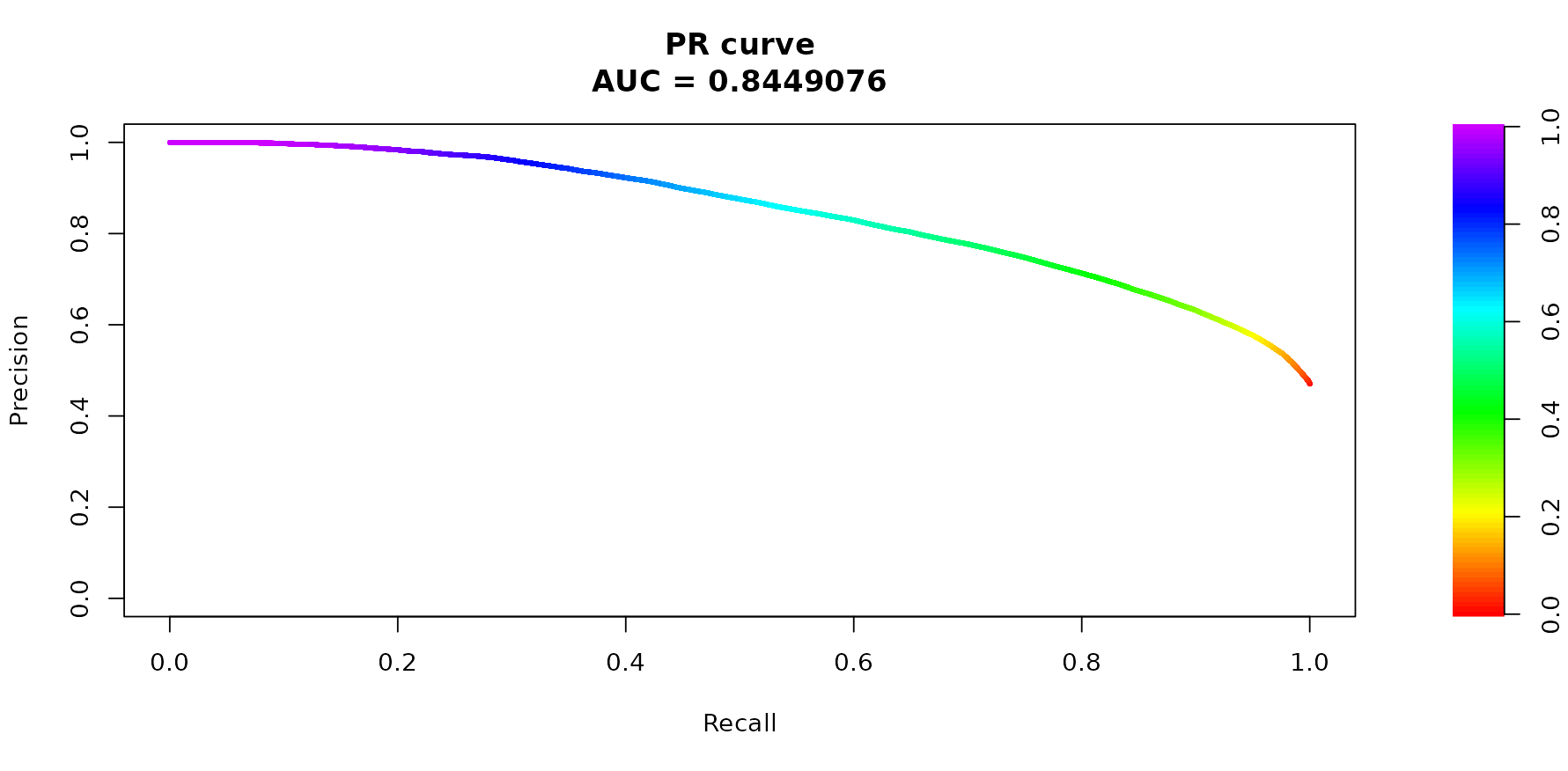}
  \caption{Evaluation of the SOG zero or positive SOG classification model using out-of-fold predictions shows the precision recall curve, with color indicating the decision
threshold and illustrating the trade off between precision and recall across
operating points.}
  \label{fig:sog_pr}
\end{figure}

Classification performance at a probability threshold of  $\hat{p}_{\text{pos\_SOG}} \ge 0.5$
was summarized using a confusion matrix, treating positive SOG observations ($Y_i = 1$)
as the positive class. Overall accuracy
($0.760$), substantially exceeded the no information rate ($0.531$). Sensitivity for detecting $SOG >0$ was $0.703$, while specificity for identifying $SOG=0$ was $0.810$, yielding a balanced accuracy of $0.757$. Cohen’s kappa of $0.516$ indicates moderate agreement beyond chance, which is expected in the presence of class imbalance, even when overall discrimination is strong \citep{FeinsteinCicchetti1990}. Consistent with this, the high AUC indicates effective separation between zero and positive speed observations.
Precision--recall analysis yielded a PR--AUC of $0.845$, indicating strong performance in identifying positive SOG vessels while maintaining a low false positive rate (Figure~\ref{fig:sog_pr}). This substantially exceeded the baseline PR--AUC expected from the prevalence of positive SOG observations alone.

\begin{table}[h]
\normalsize
\centering
\caption{\normalsize Performance metrics for zero or positive SOG classification (positive class = Positive SOG).}
\label{tab:sog_metrics}
\begin{tabular}{lc}
\toprule
\textbf{Metric} & \textbf{Value} \\
\midrule
Accuracy                  & 0.760 \\
Balanced Accuracy         & 0.757 \\
Sensitivity (Recall)      & 0.703 \\
Specificity               & 0.810 \\
Precision (PPV)           & 0.767 \\
Negative Predictive Value & 0.754 \\
Cohen’s Kappa             & 0.516 \\
\midrule
Prevalence (Positive SOG)       & 0.471 \\
Detection Rate            & 0.331 \\
Detection Prevalence      & 0.431 \\
\bottomrule
\end{tabular}
\end{table}

% \begin{figure}[]
%   \centering
%   \includegraphics[width=.7\linewidth]{ClassificationFigures/CalibPlot_CL.png}
%   \caption{OOF Calibration curve, where binned predicted
% positive SOG probabilities are compared to observed positive SOG frequencies; the dashed
% line denotes perfect calibration. Point size reflects the number of observations
% per probability bin. Together, these diagnostics demonstrate strong discrimination
% and well calibrated probabilistic predictions. The confusion matrix at the
% selected threshold ($\hat{p}_{\mathrm{pos\_SOG}} \ge 0.5$).
% }
%   \label{fig:sog_cal}
% \end{figure}
\subsection{Evaluation of Regression}

Cross-validation was performed at the level of individual AIS records. Random effects for MMSI, cell, and TimeID were estimated using training data only and were not shared across folds, ensuring that validation predictions were generated without access to validation outcomes. Predictive performance was evaluated using five fold cross-validation, with the GPBoost model trained on a square root transformation of vessel speed ($\sqrt{\mathrm{SOG}}$) and out-of-fold predictions generated for each fold. Performance was stable across folds, with $R^2_{\mathrm{SOG}} \approx 0.77$ and absolute errors near 1~knot (MAE $\approx 1.04$~kn; RMSE $\approx 1.77$~kn).

\begin{figure}[h]
\centering
\includegraphics[width=.5\linewidth]{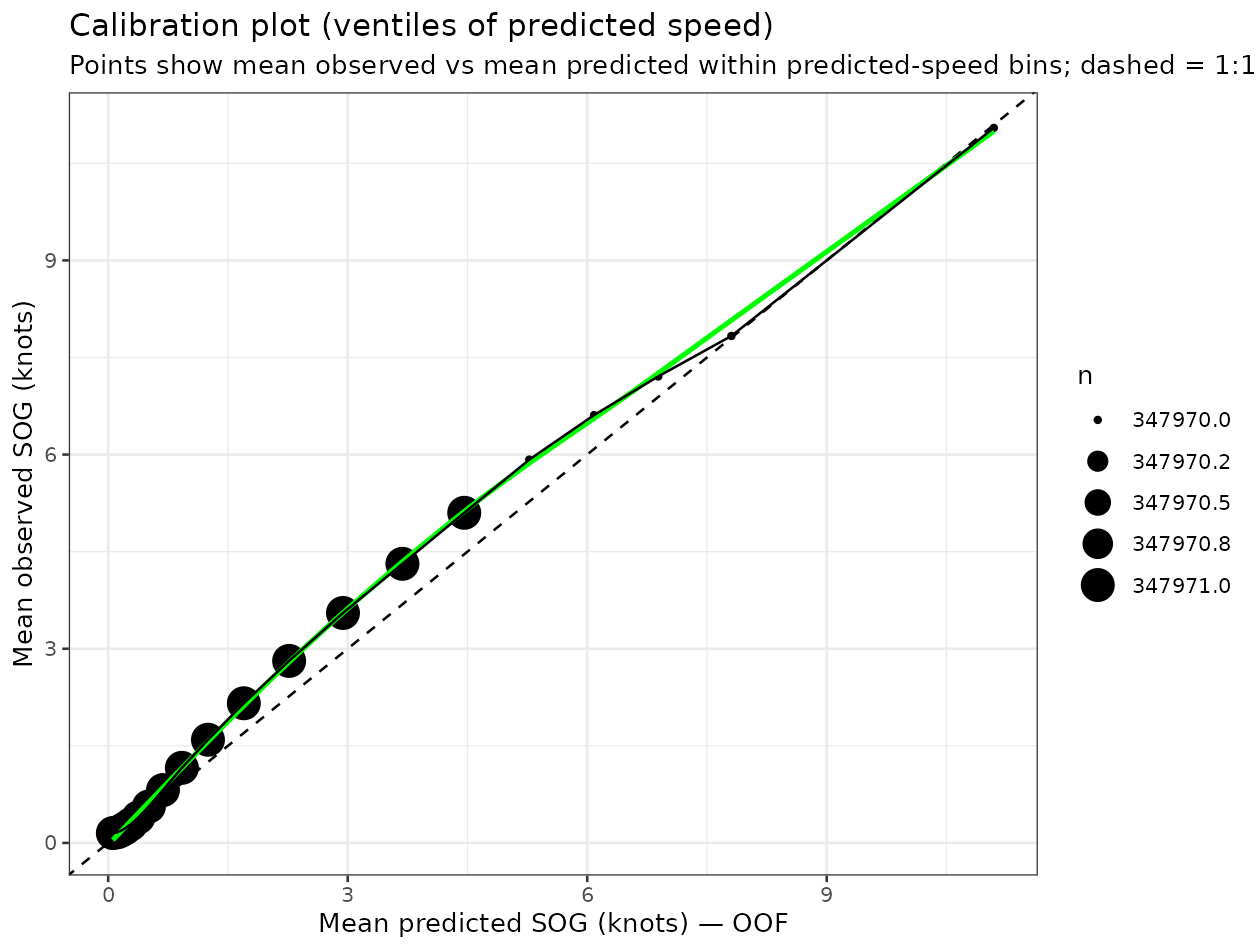}
\caption{Out-of-fold calibration plot for the GPBoost model. Mean observed and predicted vessel speeds are compared across equal frequency bins of predicted SOG. The dashed line shows perfect calibration, and the smooth curve highlights systematic deviations.}
\label{fig:calibration_sog}
\end{figure}

Calibration was assessed using out-of-fold predictions by comparing mean observed and predicted speeds within equal frequency bins (Figure~\ref{fig:calibration_sog}). The calibration curve closely follows the 1:1 line, indicating minimal bias.

Based on out-of-fold diagnostic summaries, absolute prediction error increased gradually with vessel speed, with mean absolute error remaining below approximately 1.5 knots for the majority of observations ($\mathrm{SOG} \leq 6$ knots) and increasing at higher speeds where data were comparatively sparse. The close agreement between mean and median errors indicates robustness to outliers, while the gradual increase in upper tail error suggests that uncertainty at higher speeds reflects limited data availability rather than systematic model bias.

\begin{table}[h]
\footnotesize
\centering
\caption{Out-of-fold predictive performance of the GPBoost speed model. Metrics are reported on both the transformed scale (training scale) and the original SOG (SOG) scale.}
\label{tab:gpboost_oof_metrics}
\begin{tabular}{lccc}
\toprule
\textbf{Scale} & \textbf{$R^2$} & \textbf{MAE} & \textbf{RMSE} \\
\midrule
Square Root Transform & 0.798 & 0.331 & 0.475 \\
Original SOG scale & 0.770 & 1.035 & 1.767 \\
\bottomrule
\end{tabular}
\end{table}

% Minor deviations at very low predicted speeds are expected due to limitations of AIS derived speed estimates, where GPS positional uncertainty can be comparable to or larger than true vessel displacement near stationarity, leading to increased variability in reported speeds \citep{harati2007analysis,last2015analysis}.

% \begin{figure}[]
% \centering
% \includegraphics[width=.5\linewidth]{RegressionFigure/PredError_RG.png}
% \caption{Out-of-fold prediction error on the original SOG scale (knots) stratified by observed speed bins. Panels report mean absolute error (MAE), median absolute error, root mean squared error (RMSE), and the 90th percentile of absolute error (P90AE). Error magnitudes increase smoothly with vessel speed, reflecting physical scaling and reduced data density at higher speeds rather than systematic model bias. Relative error remains stable across speed regimes.}
% \label{fig:error_by_sog}
% \end{figure}

\section{Results}

\subsection{Binary Classification Results}

\begin{figure}[H]
    \centering
    \includegraphics[width=.7\linewidth]{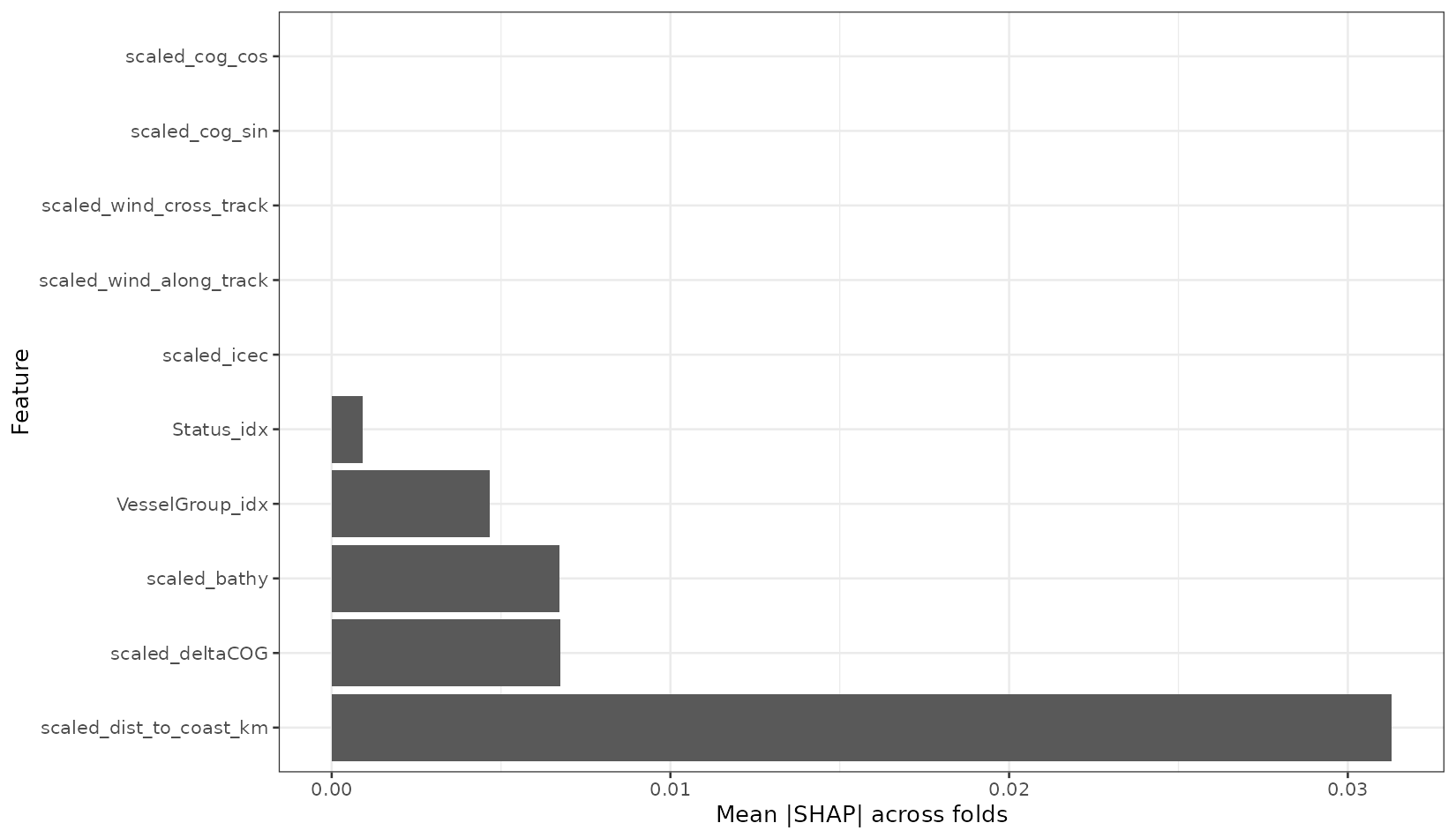}
    \caption{\normalsize Global mean absolute SHAP values for the zero or positive SOG classification model. Larger values indicate greater average influence on the probability of positive SOG.}
    \label{fig:shap-global}
\end{figure}

Figure~\ref{fig:shap-global} summarizes the global importance of predictors in the classification model based on mean absolute SHAP values computed from out-of-fold predictions. Distance to coast was the most influential predictor, indicating that proximity to shore strongly determined whether a vessel was classified as having zero or positive SOG.
Change in course over ground ($\Delta$COG) and bathymetric depth ranked as the second and third most influential predictors, respectively. Here, $\Delta$COG functioned as a steering indicator: small COG changes were associated with a higher probability of positive SOG, whereas large or frequent heading changes reduced the probability of positive SOG. Bathymetric depth contributed additional explanatory power, with vessels more likely to exhibit positive SOG in deeper waters than in shallow coastal regions.
Vessel group and navigational status exhibited systematic effects on the probability of positive SOG, but their global contributions were modest relative to distance to coast and $\Delta$COG. Wind along track, wind cross track, and sea ice concentration showed comparatively small mean absolute SHAP values, indicating limited average influence on zero versus positive SOG classification. Sine and cosine COG similarly showed low importance, suggesting that absolute heading direction played a minor role once proximity to shore and maneuvering intensity were accounted for. 

% Figure~\ref{fig:Vessel_Shap and Figure} presents SHAP based feature effects stratified by vessel group (panela) and navigational status (panelb), highlighting variation in the probability of positive SOG across operational contexts. Points represent mean SHAP contributions for each category, with point size reflecting the number of observations contributing to each estimate.

\begin{figure}[H]
         \centering
        \includegraphics[width=.55\linewidth]{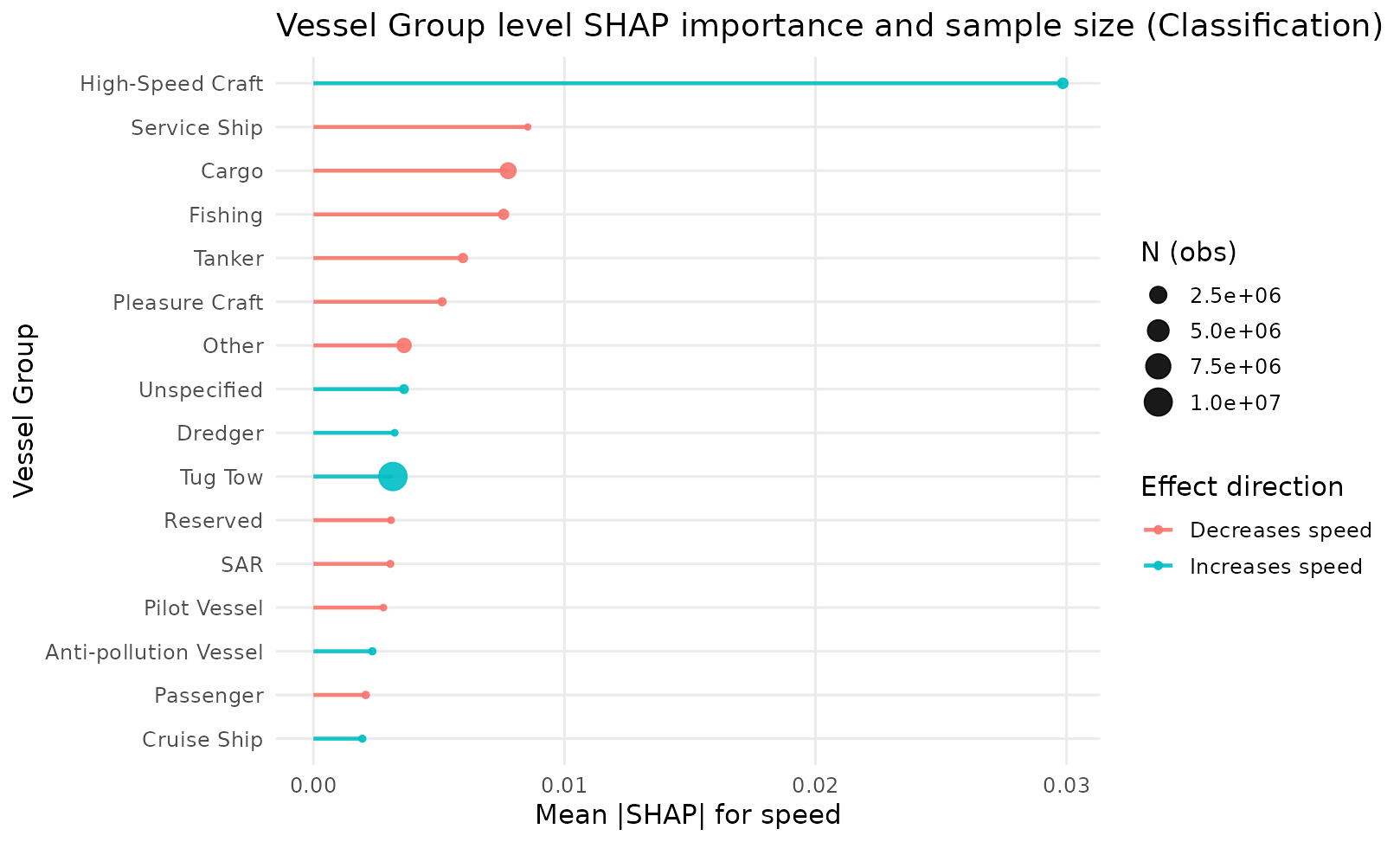}
        \caption{SHAP summary for the SOG > 0 classification model, showing the contribution of vessel group to positive SOG predictions.}
        \label{fig:Vessel_Shap}
    \end{figure}

Across vessel groups (Figure~\ref{fig:Vessel_Shap}), high speed craft exhibited the highest probability of positive SOG. Tug-Tow vessels were also more likely to be classified as having positive SOG than zero SOG. Dredgers, cruise ships, and unspecified vessel categories showed weak positive deviations from the model baseline, indicating mixed zero and positive SOG observations. In contrast, cargo vessels, tankers, fishing vessels, and service ships exhibited negative mean SHAP values, indicating that they were more frequently associated with zero SOG than positive SOG relative to the baseline. Pleasure craft and vessels in the ``other’’ category displayed similar negative contributions, reflecting intermittent positive SOG or prolonged periods of zero SOG.

\begin{figure}[H]
    \centering
        \includegraphics[width=.55\linewidth]{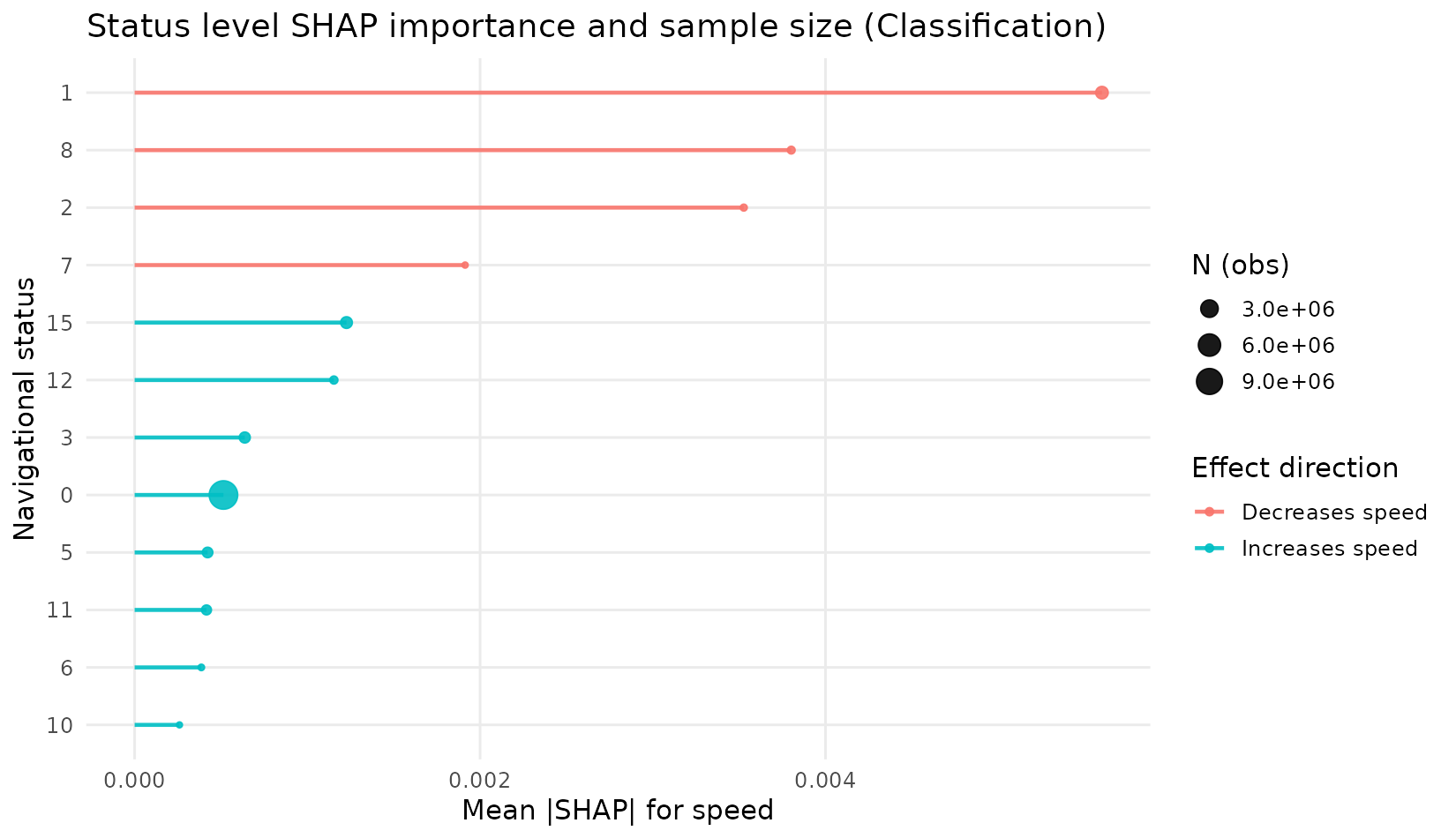}
        \caption{SHAP summary for the SOG > 0 classification model, showing the contribution of navigational status to positive SOG.}
        \label{fig:Status_Shap}
    \end{figure}

Navigational status further differentiated zero versus positive SOG outcomes (Figure~\ref{fig:Status_Shap}). Status~1 (at anchor) showed the strongest negative SHAP value, indicating an almost certain classification as zero SOG. Status~8 (underway sailing) and Status~2 (not under command) were also associated with reduced probability of positive SOG. In contrast, Status~0 (underway using engine) exhibited the strongest positive contribution, corresponding to a high likelihood of positive SOG. Statuses associated with active steering or transit such as Status~3 (restricted maneuverability), Status~12 (pushing ahead/towing alongside), and Status~15 (undefined/default) also contributed positively to the classification of positive SOG.
    % Taken together, the SHAP results show that zero versus positive SOG classification is driven primarily by operating context and short term navigational dynamics. Distance to coast and bathymetric depth distinguish nearshore, shelf, and offshore settings associated with different activity patterns, while changes in course over ground capture steering intensity that separates sustained transit from localized turning, loitering, or holding position.

% Vessel group and navigational status further modify movement probability by shifting baseline behavior in ways that are consistent with known operational roles. Certain vessel types and status codes are systematically more likely to be stationary or moving, but these effects are secondary to the dominant influence of where the vessel is operating and how it is maneuvering.

\subsubsection{Marginal Effects of Key Covariates}
\begin{figure}[htbp]
    \centering

    \begin{subfigure}{0.5\linewidth}
        \centering
        \includegraphics[width=\linewidth]{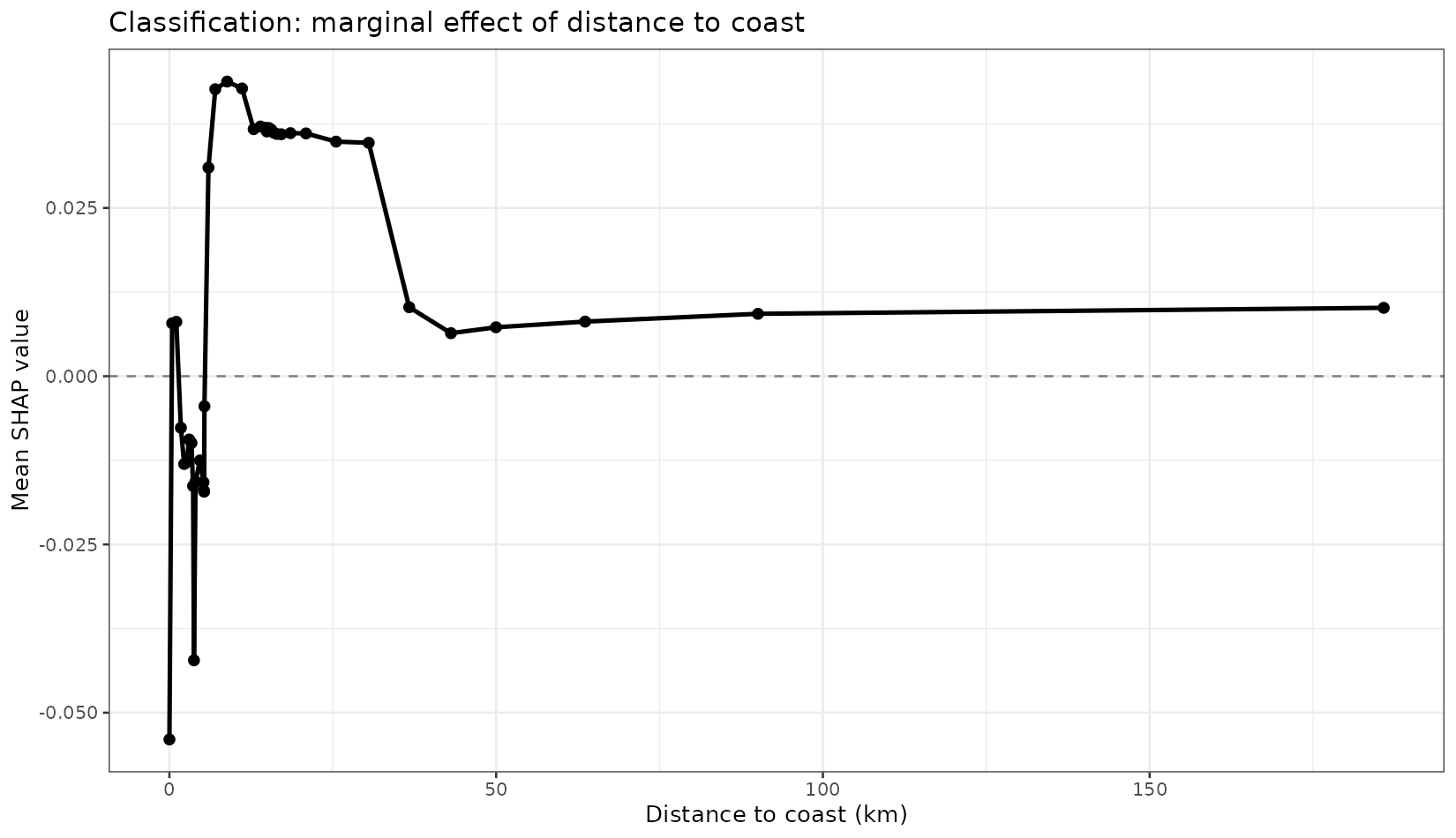}
        \caption{Distance to coast}
        \label{fig:shap_dist_coast}
    \end{subfigure}\hfill
    \begin{subfigure}{0.5\linewidth}
        \centering
        \includegraphics[width=\linewidth]{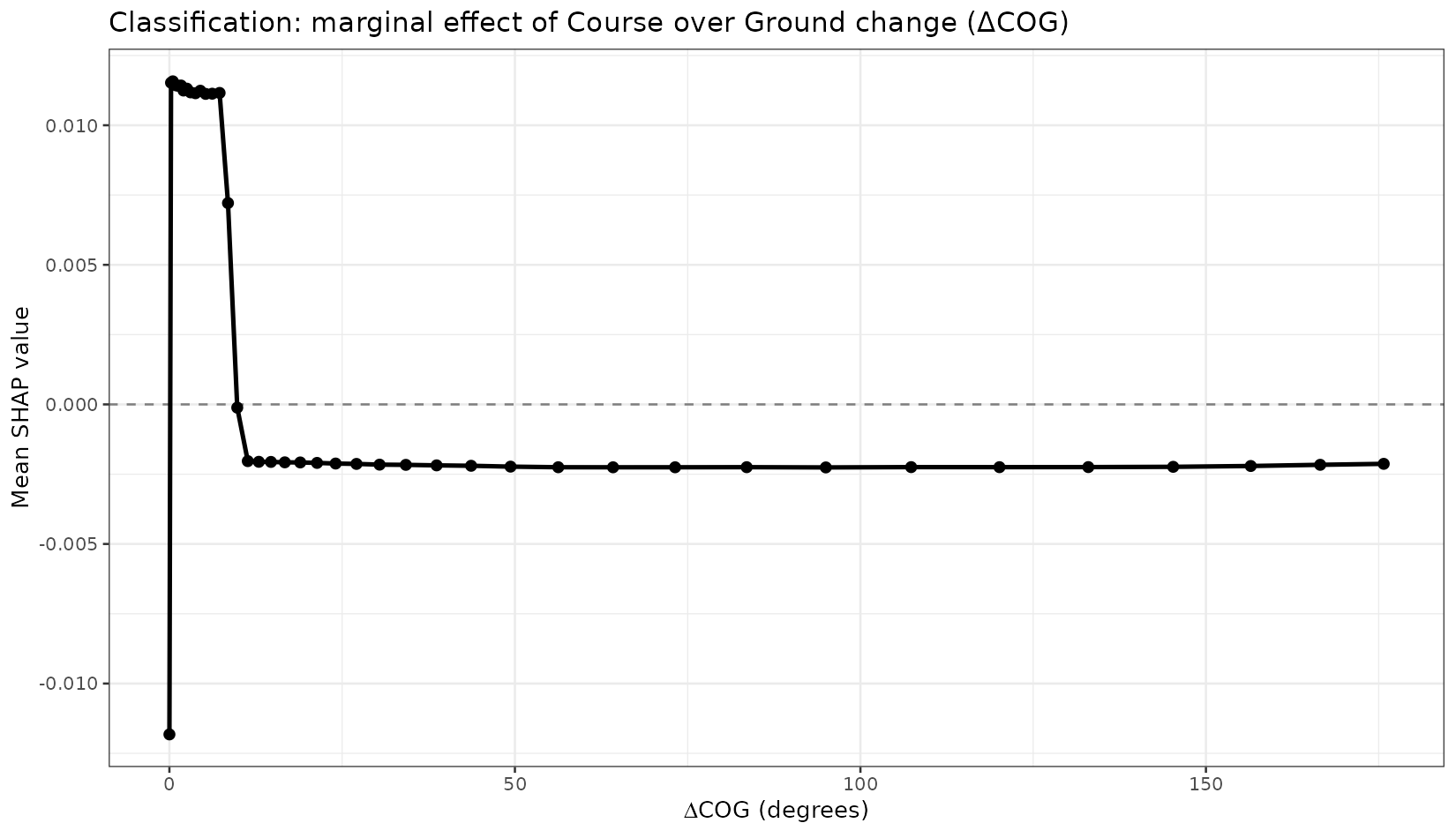}
        \caption{$\Delta$COG}
        \label{fig:shap_deltaCOG}
    \end{subfigure}

    \vspace{0.6em}

    \begin{subfigure}{0.5\linewidth}
        \centering
        \includegraphics[width=\linewidth]{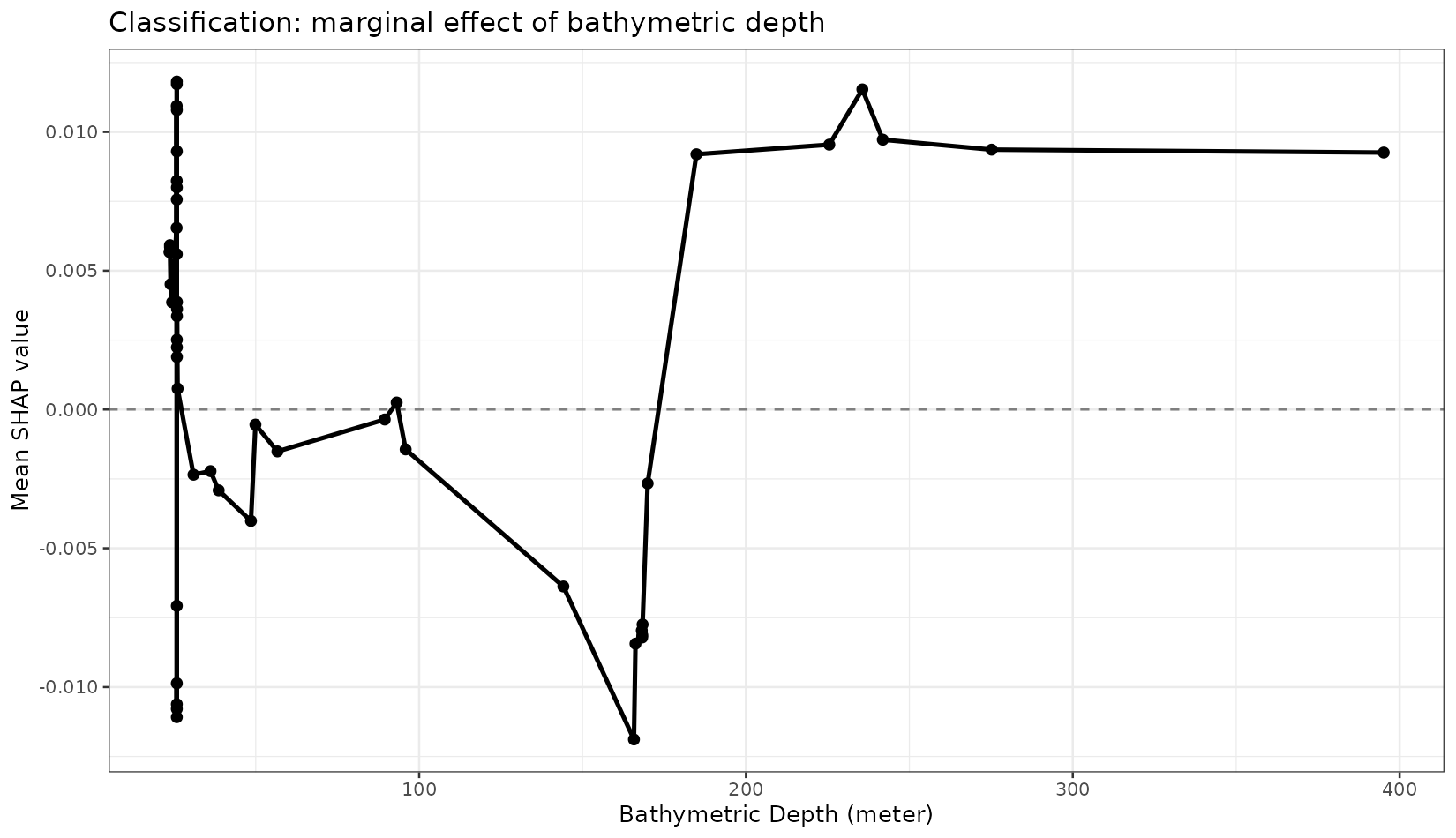}
        \caption{Bathymetric depth}
        \label{fig:shap_bathy}
    \end{subfigure}

    \caption{\normalsize
    Marginal SHAP effects for key predictors in the SOG $>$ 0 classification model.
    Panels show mean SHAP values on the probit scale across binned covariates for
    (a) distance to coast,
    (b) change in course over ground ($\Delta$COG), and
    (c) bathymetric depth.
    Positive SHAP values indicate an increased probability of positive vessel speed.
    }
\end{figure}

SHAP values were reported on the latent probit scale, corresponding to the model’s linear predictor prior to transformation into probabilities. Distance to coast (Figure~\ref{fig:shap_dist_coast}) showed a strong nonlinear marginal effect on positive SOG probability. SHAP values were strongly negative within 2–3~km of shore, increased sharply between 5 and 30~km, and stabilized near zero to slightly positive beyond 40–50~km offshore, indicating diminishing influence in fully offshore environments.
Change in course over ground ($\Delta$COG; Figure~\ref{fig:shap_deltaCOG}) showed a clear threshold effect on positive SOG probability. SHAP values were strongly positive for small heading changes ($\Delta$COG $<5^\circ$), declined rapidly between $5^\circ$ and $10^\circ$, and became persistently negative beyond $10^\circ$, indicating reduced likelihood of positive SOG during increased steering.
Bathymetric depth (Figure~\ref{fig:shap_bathy}) showed a nonlinear effect on positive SOG probability, with near-zero to negative SHAP values in shallow waters, a pronounced minimum at intermediate depths, and a sharp increase to positive values beyond 180–200~m, indicating higher likelihood of positive SOG in deeper offshore waters.
% \begin{figure}[h]
%     \centering
%     \includegraphics[width=.6\linewidth]{ClassificationFigures/ME_disttocoast.png}
%     \caption{\normalsize 
% Marginal SHAP effect of distance to coast on binary classification. Mean SHAP values (probit scale) are shown across distance-to-coast bins. Positive values indicate increased probability of positive SOG. The probability rises sharply in near coastal zones ($\approx 5-30$ km) and stabilizes farther offshore.}
%     \label{fig:shap_dist_coast}
% \end{figure}
% \begin{figure}[h]
%     \centering
%     \includegraphics[width=.6\linewidth]{ClassificationFigures/ME_deltaCOG.png}
%     \caption{\normalsize 
% Marginal SHAP effect of $\Delta COG$ on SOG classification. Mean SHAP values (probit scale) indicate high positive SOG probability for small heading changes and reduced probability as $\Delta COG$ increases.}
%     \label{fig:shap_deltaCOG}
% \end{figure}

% \begin{figure}[h]
%     \centering
%     \includegraphics[width=.6\linewidth]{ClassificationFigures/ME_wmb.png}
%     \caption{\normalsize 
% Marginal SHAP effect of bathymetric depth on SOG classification. Mean SHAP values (probit scale) across depth bins indicate reduced positive SOG probability in shallow waters and a sharp increase beyond 180–200 m.}
%     \label{fig:shap_bathy}
% \end{figure}

% Taken together, these marginal effects show that zero versus positive SOG classification depends primarily on spatial setting and short term changes in vessel COG. Distance to coast and bathymetric depth distinguish nearshore, shelf, and offshore locations associated with different SOG outcomes.
\subsection{Positive Speed Regression Results}
 Vessel SOG regression model results showed that distance to coast emerged as the dominant predictor, indicating that spatial proximity to shore was the strongest determinant of vessel speed once vessels were underway. Change in course over ground ($\Delta$COG) and bathymetric depth ranked second and third, respectively, highlighting the importance of navigational stability and seafloor constraints in shaping achievable speed.
Directional components of COG, represented by sine and cosine encodings, exhibited moderate importance, indicating persistent directional effects on vessel speed while avoiding artificial discontinuities associated with angular variables. Their contribution was secondary to $\Delta$COG. Operational attributes, including navigational status and vessel group, contributed additional explanatory power but were less influential than the primary spatial covariates mentioned above.
Environmental variables wind along track, wind cross track, and sea ice concentration showed comparatively low mean absolute SHAP values. 

\begin{figure}[h]
    \centering
    \includegraphics[width=0.7\linewidth]{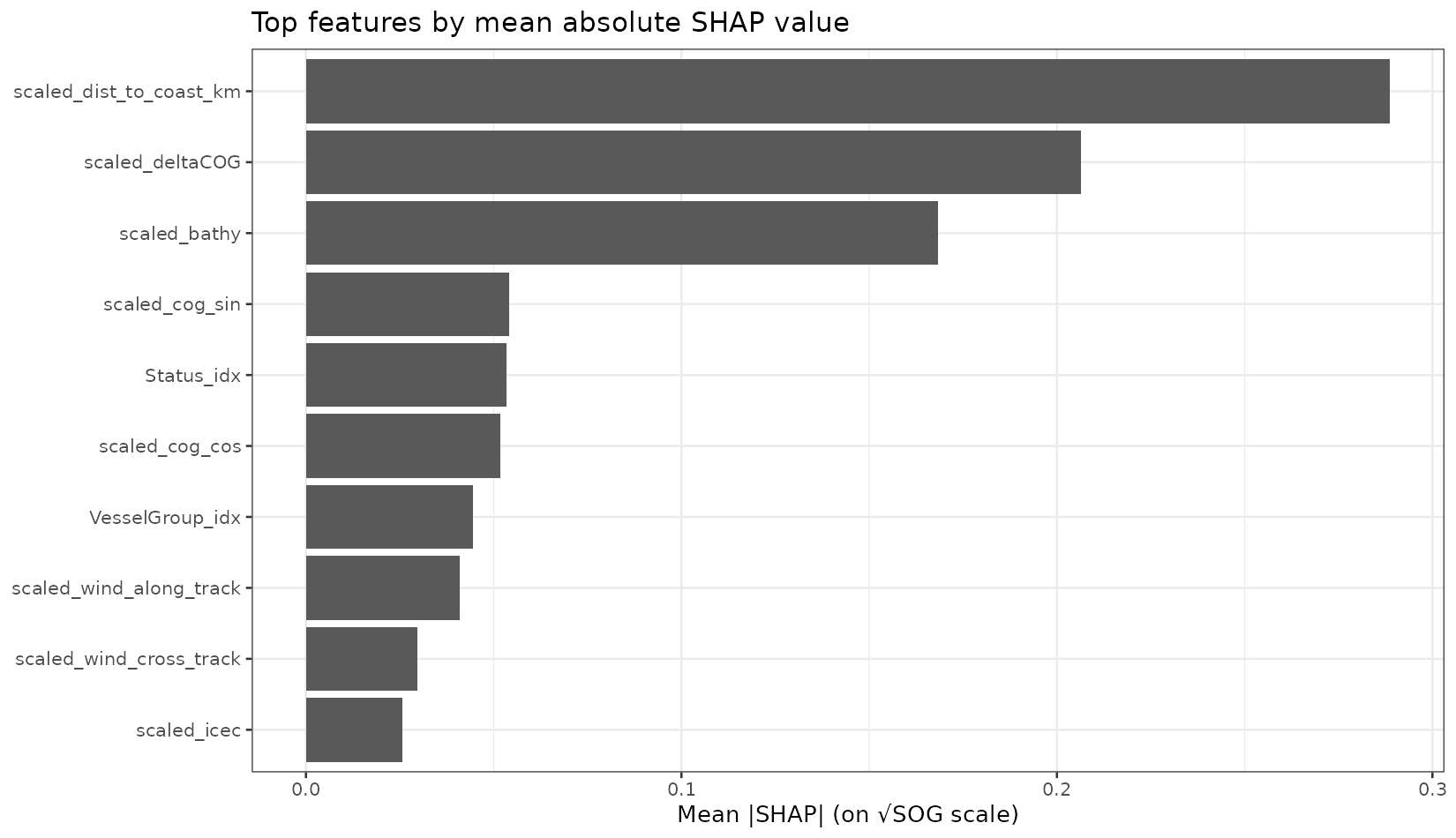}
    \caption{\normalsize Global feature importance for the vessel speed regression model.}
    \label{fig:mean_shap}
\end{figure}

\subsubsection{Covariate Marignal Effects on Vessel Speed}

\begin{figure}[H]
    \centering

    \begin{subfigure}{0.48\linewidth}
        \centering
        \includegraphics[width=\linewidth]{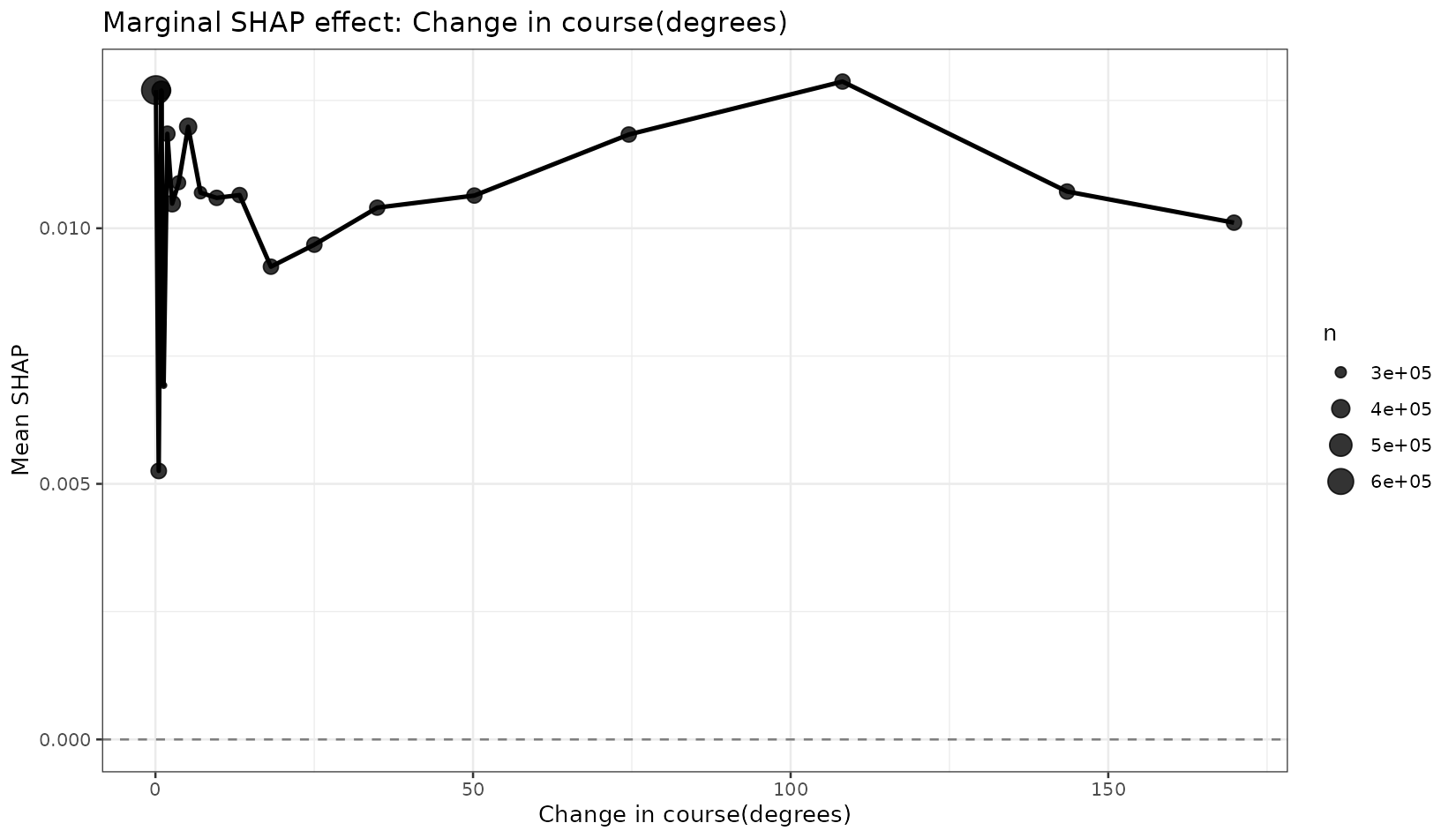}
        \caption{$\Delta$COG}
        \label{fig:reg_deltaCOG}
    \end{subfigure}\hfill
    \begin{subfigure}{0.5\linewidth}
        \centering
        \includegraphics[width=\linewidth]{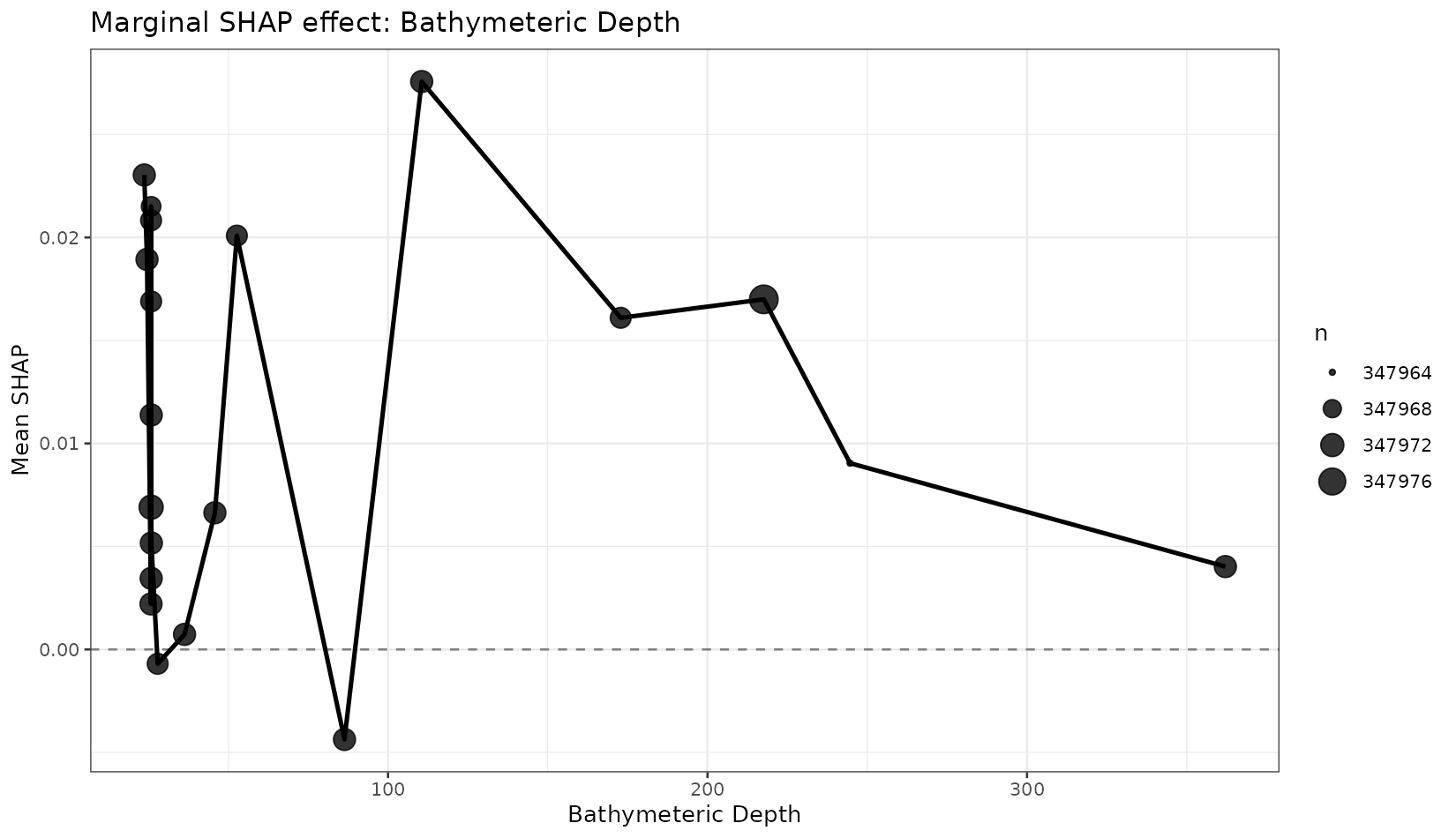}
        \caption{Bathymetric depth}
        \label{fig:reg_bathy}
    \end{subfigure}

    \vspace{0.6em}

    \begin{subfigure}{0.5\linewidth}
        \centering
        \includegraphics[width=\linewidth]{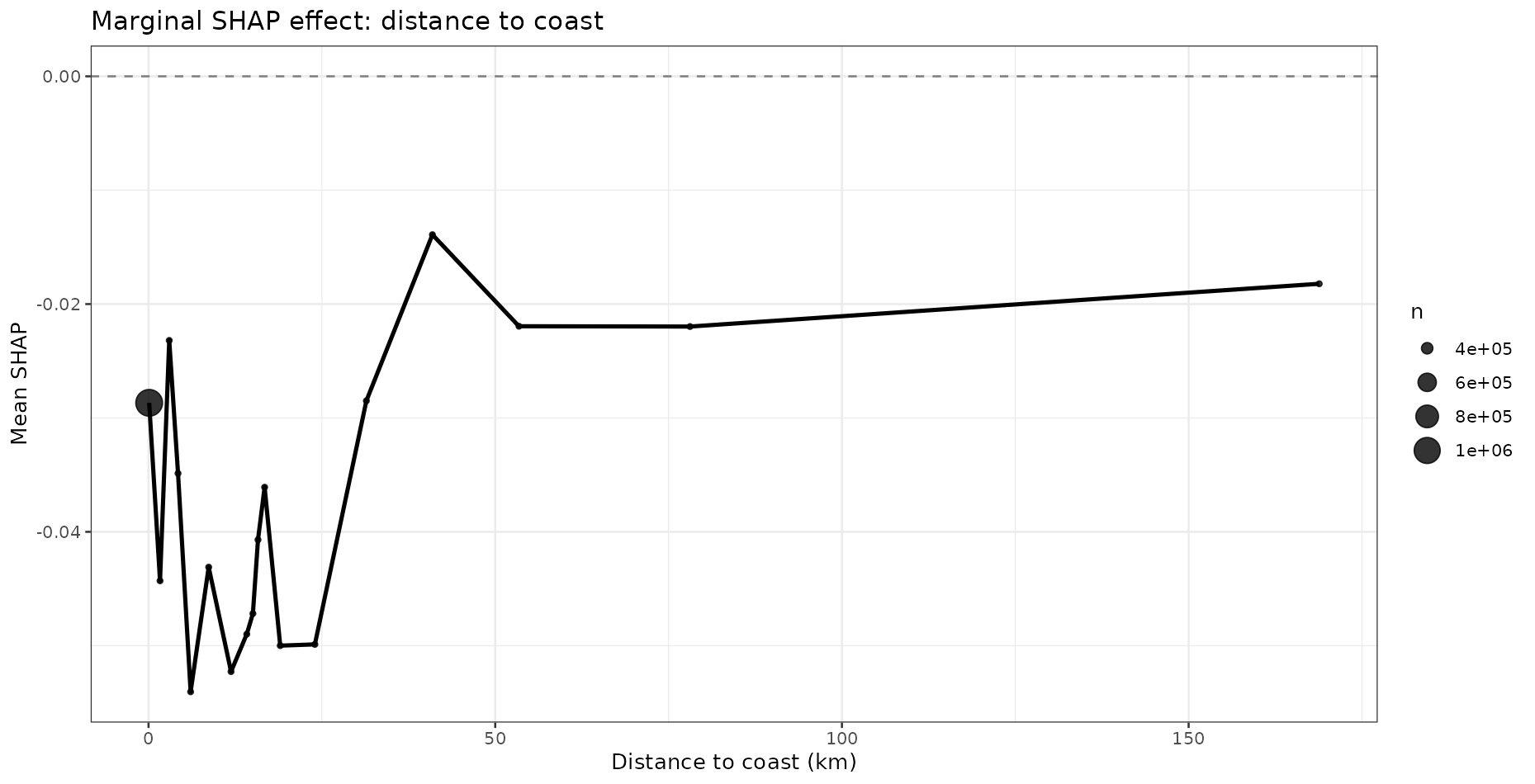}
        \caption{Distance to coast}
        \label{fig:reg_dist_coast}
    \end{subfigure}\hfill
    \begin{subfigure}{0.5\linewidth}
        \centering
        \includegraphics[width=\linewidth]{RegressionFigure/distocoast_SHAP.png}
        \caption{Distance to coast (alternate binning)}
        \label{fig:reg_dist_coast_alt}
    \end{subfigure}

    \caption{\normalsize
    Marginal SHAP effects for key spatial and navigational predictors in the SOG $>$ 0 regression model.
    Panels show mean SHAP values (on the $\sqrt{\mathrm{SOG}}$ scale) across binned covariates for
    (a) change in course over ground ($\Delta$COG),
    (b) bathymetric depth,
    (c) distance to coast, and
    (d) an alternative binning of distance to coast.
    Marker size reflects bin sample size.
    }
\end{figure}

Distance to coast (Figure~\ref{fig:reg_dist_coast}) showed a strong nonlinear effect on predicted SOG, with low speeds nearshore (0 – 10~km), a rapid increase offshore (15 – 40~km), and stabilization beyond 50 – 60~km

% \begin{figure}[h]
%     \centering
%     \includegraphics[width=.8\linewidth]{RegressionFigure/distocoast_SHAP.png}
%     \caption{\normalsize Marginal SHAP effect of distance to coast on predicted vessel speed. Mean SHAP values (on the $\sqrt{\mathrm{SOG}}$ scale) across distance bins show strong speed suppression nearshore (0--10~km), a rapid increase offshore ($\approx$15 -- 40~km), and stabilization beyond $\approx$50 -- 60~km. Marker size indicates bin sample size.}
%     \label{fig:reg_dist_coast}
% \end{figure}
Bathymetric depth (Figure~\ref{fig:reg_bathy}) showed a nonlinear positive effect on predicted vessel speed, with near-zero effects in shallow waters (0 – 50~m), a sharp increase at intermediate depths (80 – 150~m), and stabilization beyond 200 – 250~m, indicating diminishing influence in deeper offshore regions.

\begin{figure}[h]
\centering
\includegraphics[width=.8\linewidth]{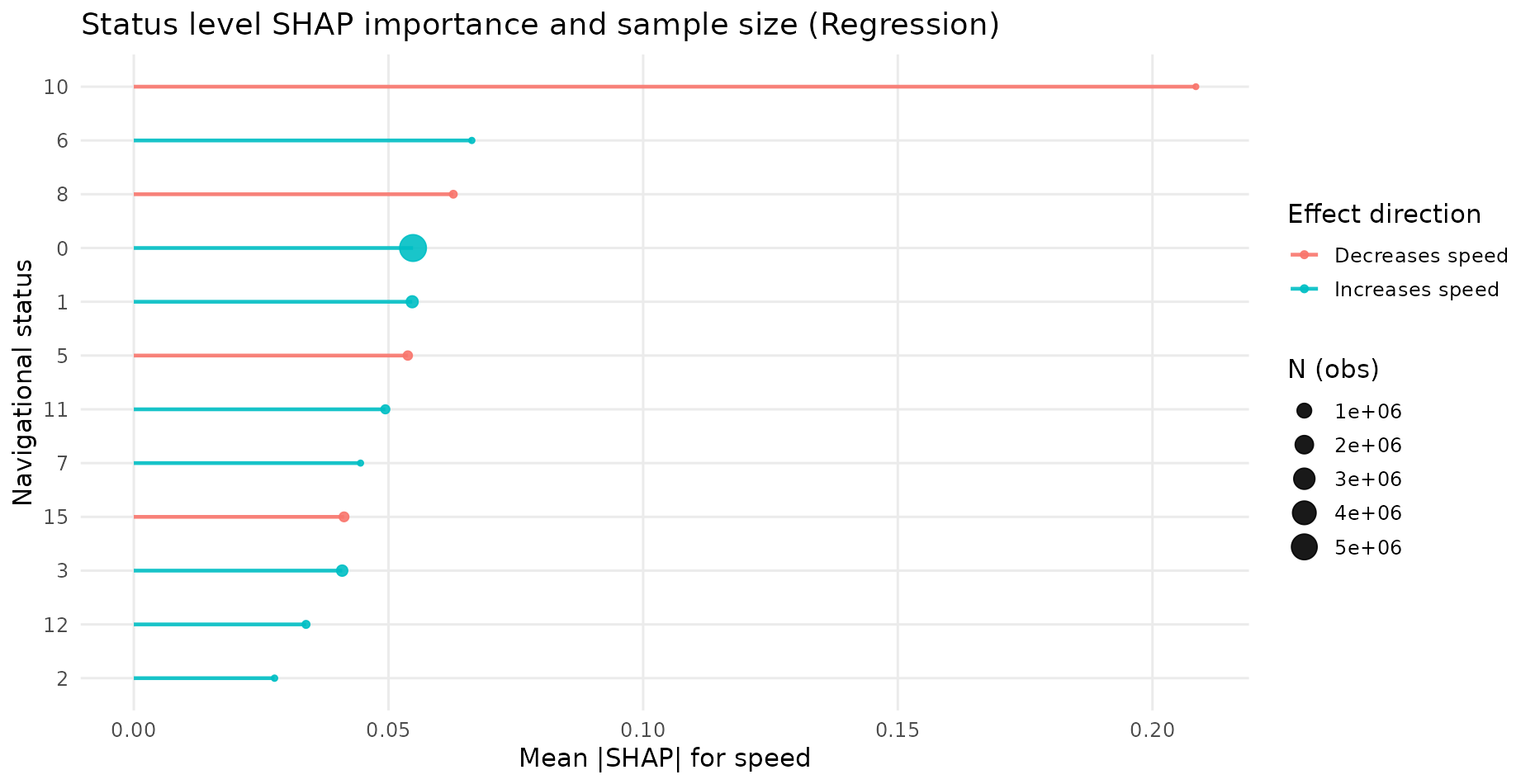}
        \caption{SHAP summary for the SOG $>$ 0 regression model, showing the effect of navigation status on predicted vessel speed.}
        \label{fig:status}
\end{figure}

Change in course over ground ($\Delta$COG; Figure~\ref{fig:reg_deltaCOG}) showed a nonlinear effect on predicted vessel speed, with highest speeds for small heading changes ($<5^\circ$), declining at moderate changes (5 – 20$^\circ$), and persistently lower speeds at larger changes, indicating reduced sustained speed during increased steering.
Sea ice concentration and wind exhibited comparatively small marginal SHAP effects and are therefore not shown. Their limited  influence reflected averaging across a wide range of operational contexts, including differences in vessel type, navigation status, and spatial setting. When other environmental, spatial, and operational variables were accounted for, ice and wind had a limited overall influence on vessel speed, although their effects may become more pronounced under specific ice or wind conditions.
Navigational status exhibited clear, discrete influences on vessel speed (Figure~\ref{fig:status}), with both the direction and magnitude of effects varying substantially across operational states.
Mean SHAP values for navigational status spanned a range from approximately $-0.20$ to $+0.06$ on the $\sqrt{\mathrm{SOG}}$ scale.
Negative SHAP values corresponded to reductions in predicted SOG relative to the model baseline, while positive values indicated increases.
Status~10 (reserved; dangerous goods/high speed craft) showed the strongest negative effect, with a mean SHAP value of approximately $-0.20$, corresponding to a substantial reduction in predicted SOG.
On the original SOG scale, this magnitude implied a meaningful decrease in SOG relative to baseline operating conditions.
Other restrictive states, including Status~15 (not defined) and Status~5 (moored), exhibited smaller but consistent SOG reductions, with mean SHAP values on the order of $-0.04$ to $-0.06$.
% \begin{figure}[h]
%     \centering
%     \includegraphics[width=.8\linewidth]{RegressionFigure/bathy_SHAP.png}
%     \caption{\normalsize Marginal SHAP effect of bathymetric depth on predicted vessel speed. Mean SHAP values (on the $\sqrt{\mathrm{SOG}}$ scale) across depth bins indicate lowest speeds in shallow waters ($<50$~m), a sharp increase at intermediate depths (80 -- 150~m), and stabilization beyond $\approx$200 -- 250~m. Marker size reflects bin sample size.}
%     \label{fig:reg_bathy}
% \end{figure}

% \begin{figure}[h]
%     \centering
%     \includegraphics[width=.8\linewidth]{RegressionFigure/deltaCOG_SHAP.png}
%     \caption{\normalsize Marginal SHAP effect of change in course over ground ($\Delta\mathrm{COG}$) on predicted vessel speed. Mean SHAP values (on the $\sqrt{\mathrm{SOG}}$ scale) across $\Delta\mathrm{COG}$ bins indicate highest speeds for small heading changes ($\Delta\mathrm{COG}<5^\circ$) and decreasing speeds with increasing maneuvering. Marker size reflects bin sample size.}
%     \label{fig:reg_deltaCOG}
% \end{figure}

In contrast, Status~6 (aground), Status~7 (engaged in fishing), and Status~12 (pushing ahead/towing alongside) displayed positive mean SHAP values of approximately $+0.04$ to $+0.06$, indicating increased predicted SOG conditional on positive SOG.
These positive contributions suggested that, when vessels were actively navigating under these statuses, higher SOG were more frequently observed relative to the baseline.
By comparison, common navigational states such as underway using engine (Status~0), at anchor (Status~1), and restricted maneuverability (Status~3) clustered tightly around zero, with mean SHAP magnitudes generally below $0.02$, indicating minimal marginal influence on SOG once other covariates were accounted for.

\begin{figure}[h]
 \centering     \includegraphics[width=.8\linewidth]{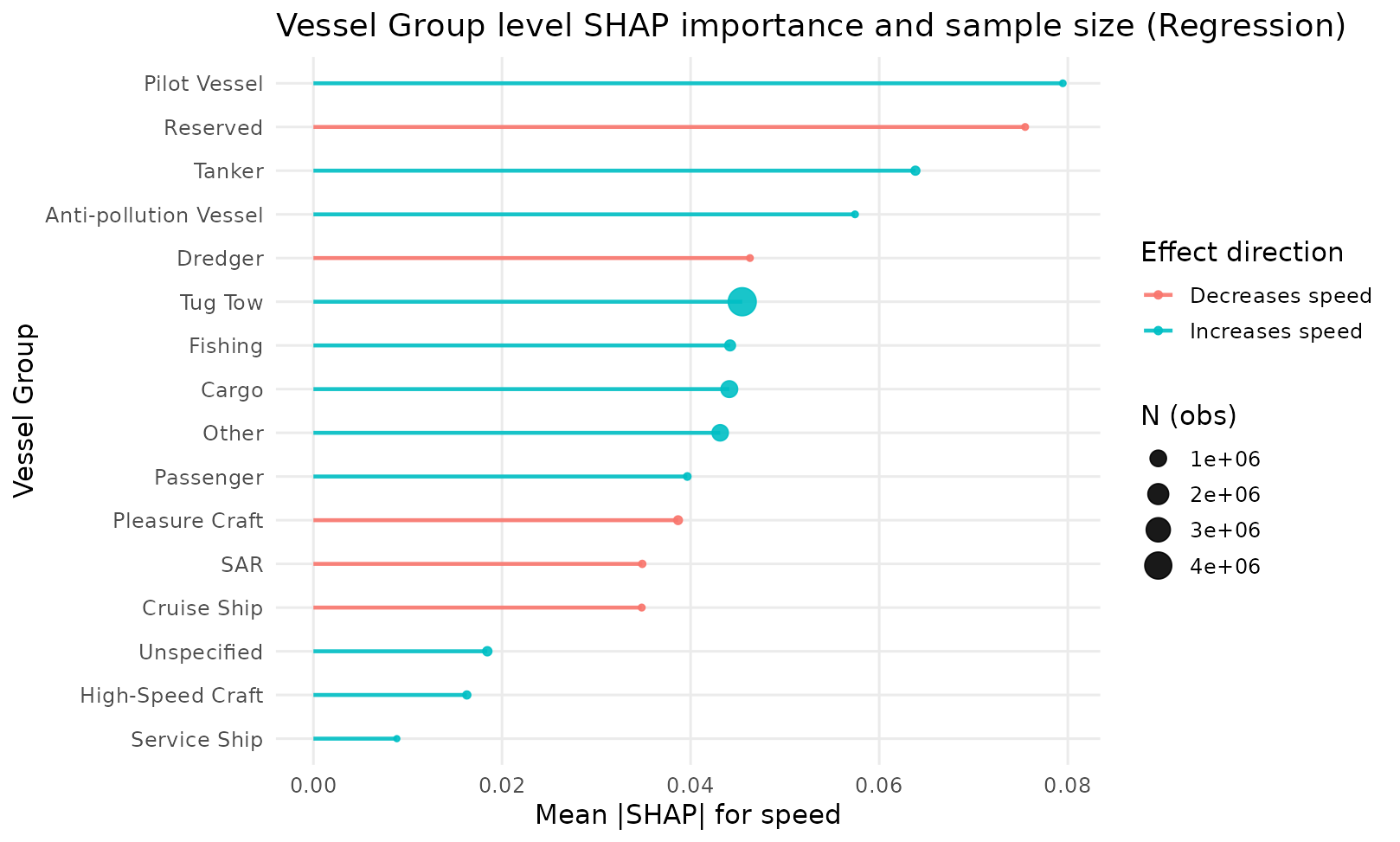}
        \caption{SHAP summary for the SOG $>$ 0 regression model, showing the effect of vessel group on predicted vessel speed.}
        \label{fig:vslgrp}
\end{figure}

Figure~\ref{fig:vslgrp} showed systematic differences in baseline vessel SOG across vessel groups, with mean SHAP values ranging approximately from $-0.06$ to $+0.18$ on the $\sqrt{\mathrm{SOG}}$ scale.
Positive SHAP values indicated vessel groups that tended to operate at higher SOG relative to the model baseline, while negative values indicated systematically lower SOG.

Passenger vessels and tankers exhibited the largest positive contributions, with mean SHAP values of approximately $+0.15$ to $+0.18$, corresponding to substantial increases in predicted SOG.
Pilot vessels and tug-tow operations also showed strong positive effects, with mean SHAP values on the order of $+0.10$ to $+0.15$.

Cargo, fishing, and other general purpose vessels displayed more moderate SOG increases, with mean SHAP values typically between $+0.05$ and $+0.08$.
In contrast, cruise ships, dredgers, anti-pollution vessels, and pleasure craft were associated with negative mean SHAP values, generally ranging from $-0.03$ to $-0.06$, indicating systematic reductions in vessel SOG.
Overall, vessel group imposed a strong structural constraint on achievable SOG, with effect magnitudes that exceeded those of individual environmental covariates and persisted across a wide range of operating conditions.
\section{Discussion}

By explicitly distinguishing zero speed from positive speed regimes, this study enabled a targeted interpretation of where and when speed related risks arose in the U.S. Arctic. Nearshore regions characterized by frequent zero speed activity were associated with increased steering, congestion, and prolonged vessel presence, whereas offshore transit corridors were primarily associated with sustained higher speeds. These nearshore operating regimes often overlapped with ecologically sensitive coastal habitats and marine mammal use areas documented in prior studies \citep{Reeves2014,Hauser2018,Halliday2020}, indicating that navigational and ecological risks may co-occur in space. Together, these findings showed that speed related risk in Arctic waters was strongly context dependent.

Across both modeling stages, spatial context captured primarily by distance to coast emerged as the dominant determinant of vessel SOG. Very close to shore, the probability of positive SOG was substantially reduced, while both the likelihood and magnitude of vessel speed increased rapidly offshore before stabilizing at greater distances. This pattern was consistent with AIS based studies documenting concentrated low speed and zero speed activity in coastal operating regions and more stable transit behavior offshore \citep{Ma2024AISCoastalSpeed}. These results indicated that proximity to shore governed both whether vessels moved and how fast they traveled when underway.

Changes in course over ground ($\Delta$COG), provided a complementary distinction between operating regimes. Small heading changes were associated with sustained transit and higher speeds, whereas large or frequent directional changes corresponded to reduced speed and a lower probability of positive SOG. Similar patterns had been documented in AIS-based studies of maneuvering intensive activities, where frequent course changes reflected localized operations rather than long distance transit \citep{jmse11051093}. Together, distance to coast and $\Delta$COG distinguished corridor based navigation from maneuvering dominated regimes in a way not captured by static vessel attributes alone.

Bathymetric depth further contributed to both classification and regression outcomes. Shallow waters were associated with a higher probability of zero-speed observations, while deeper waters supported higher and more stable transit speeds. This finding aligned with prior AIS analyses showing systematic differences in vessel behavior between shallow coastal environments and deeper offshore regions, where navigational constraints were reduced and maneuvering intensity was lower \citep{Ma2024AISCoastalSpeed}.

Operational characteristics such as vessel group and navigational status introduced systematic but secondary variation in vessel speed behavior(see supplementary figure S1,S7 and S8). Status codes indicating anchoring or mooring were strongly associated with zero SOG, while transit related statuses corresponded to positive SOG, consistent with AIS definitions. Vessel group effects similarly reflected operational roles observed in the data. However, across both classification and regression models, these operational attributes did not override the dominant influence of spatial setting and steering behavior(see supplementary figure S2,S9 and S10), as reflected in global SHAP rankings where distance to coast and $\Delta$COG consistently exceeded vessel group and status in explanatory importance.

Environmental variables, including wind speed and sea ice concentration, exhibited modest average influence on vessel SOG because their effects were highly nonlinear and strongly context dependent(see Supplementary Material Figure S3,S4,S6 and S7 ). When summarized using global SHAP measures, impacts arising only under specific conditions such as high ice concentration, adverse winds, or constrained navigation were attenuated by averaging across heterogeneous vessels, routes, and predominantly open-water conditions during the July–October season. Consequently, wind and ice could exert substantial localized influence in particular operating regimes even though their global importance appeared limited. This interpretation was consistent with prior work showing that wind and ice effects became dominant only under specific operating conditions \citep{Aksenov2017Navigability,Montewka2015-xz}.

A key implication of the two-stage framework was that it separated operating regimes that were conflated in single-stage speed models. When SOG was modeled as a single continuous response, zero speed observations dominated nearshore and steering intensive contexts, biasing predicted speeds downward and obscuring transitions between stationary behavior and sustained transit. In contrast, the two-stage structure first identified whether a vessel was likely to be underway ($\mathrm{SOG}>0$) and then modeled achievable speed conditional on transit. This decomposition allowed speed suppression near the coast to be interpreted as a combination of increased stopping and reduced conditional speed rather than as a single averaged effect. As a result, the model outputs provided a clearer operational distinction between areas dominated by steady or steering and corridors where sustained transit at higher speeds was feasible.

From an operational standpoint, this distinction helped identify where speed management efforts would be most effectively focused, including reducing stop-and-go congestion in nearshore corridors versus moderating sustained transit speeds in offshore routes.

Although the two-stage framework improved interpretability, AIS-reported SOG and navigational status may have contained measurement noise, and environmental covariates mapped at daily $0.5^{\circ} \times 0.5^{\circ}$ resolution may have smoothed localized extremes. The analysis was limited to the July–October season and excluded drivers such as ocean surface currents, which can substantially influence SOG \citep{ZHOU2020107774,su12093649} and contribute to unexplained variability, particularly in coastal corridors. Future work could address these limitations by incorporating higher-resolution environmental data, extending analyses to additional seasons, and explicitly accounting for AIS measurement uncertainty.

\begingroup
\singlespacing
\setlength{\bibsep}{4pt}   % natbib/elsarticle supports this in many setups
\bibliographystyle{elsarticle-harv}
\bibliography{Citations}
\endgroup
\appendix 
\section{}
\setcounter{figure}{0}
\renewcommand{\thefigure}{A\arabic{figure}}
\begin{figure}[H]
    \centering
    \includegraphics[width=0.7\linewidth]{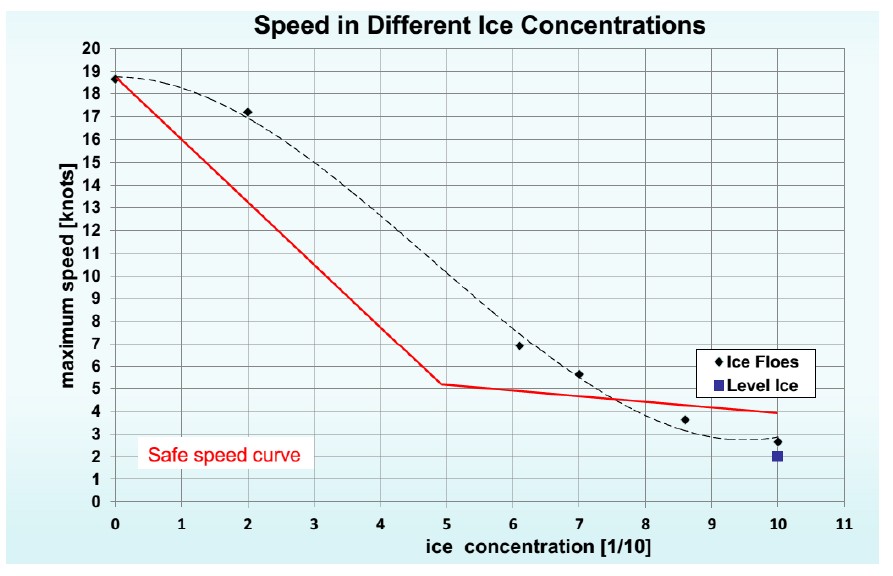}
    \caption{\normalsize
Maximum recommended vessel speed as a function of sea ice concentration. Black symbols and dashed curves show observed speed–ice relationships, while the solid red line denotes the conservative safe speed threshold used to define ice-related navigation risk. Ice concentration is expressed in tenths (0–10) and speed in knots. Adapted from ACCESS ship performance guidance \citep{HSVA2014FuelConsumption}.}
    \label{fig:Access}
\end{figure}
\end{document}

\setcounter{figure}{0}
\renewcommand{\thefigure}{S\arabic{figure}}

\section*{Supplementary Material}

This supplementary document presents additional diagnostics supporting the two-stage
GPBoost modeling framework. The diagnostics are intended to (i) evaluate model calibration
across key environmental, spatial, and operational contexts, and (ii) illustrate
context-dependent effects that are not fully captured by global importance measures.
All diagnostics are based on out-of-fold (OOF) predictions and are descriptive rather than
formal tests of parametric interaction terms.

Point sizes in all figures reflect the number of AIS observations contributing to each bin.

% ============================================================
\section{Supplementary Classification Diagnostics}
\label{supp:classification}

These diagnostics examine how environmental and operational variables influence the
probability that an AIS observation corresponds to positive vessel speed
($\mathrm{SOG} > 0$).

\subsection{Positive-speed probability as a function of distance to coast by vessel group}

\begin{figure}[H]
  \centering
  \includegraphics[width=\textwidth]{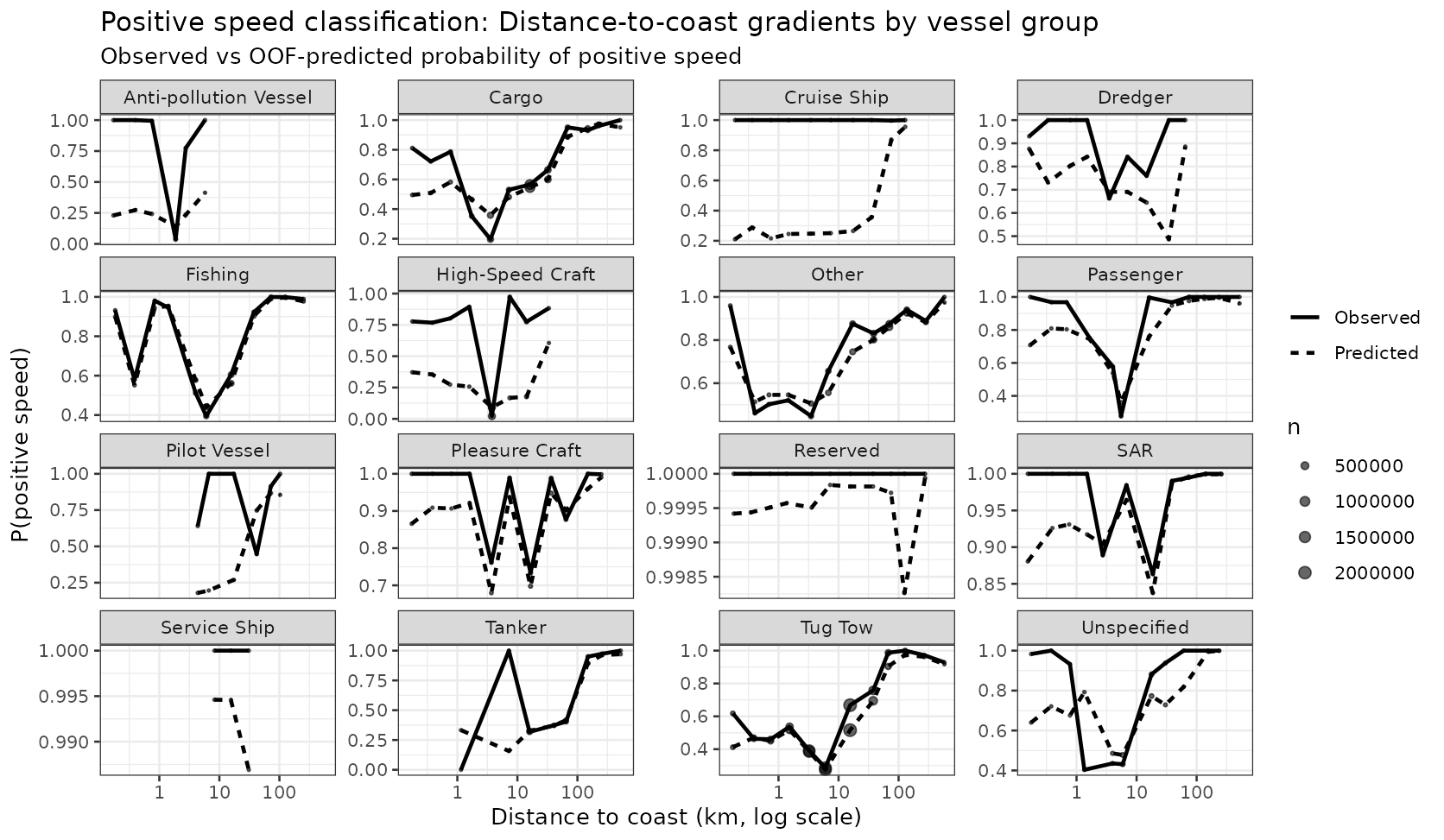}
  \caption{
  Positive-speed probability as a function of distance to coast by vessel group.
  Solid lines indicate observed probabilities; dashed lines show out-of-fold predictions.
  Distance to coast is shown on a logarithmic scale.
  }
  \label{fig:clf_dist_vesselgroup}
\end{figure}

Figure~\ref{fig:clf_dist_vesselgroup} shows observed and predicted probabilities of $SOG>0$ as a function of distance to coast, stratified by vessel group. Across nearly all
groups, probabilities of positive speed are lowest nearshore and increase rapidly offshore.

The close agreement between observed and predicted curves indicates that the classifier
captures heterogeneous operational regimes without systematic bias.

\subsection{Positive SOG probability as a function of distance to coast by navigational status}

\begin{figure}[H]
  \centering
  \includegraphics[width=\textwidth]{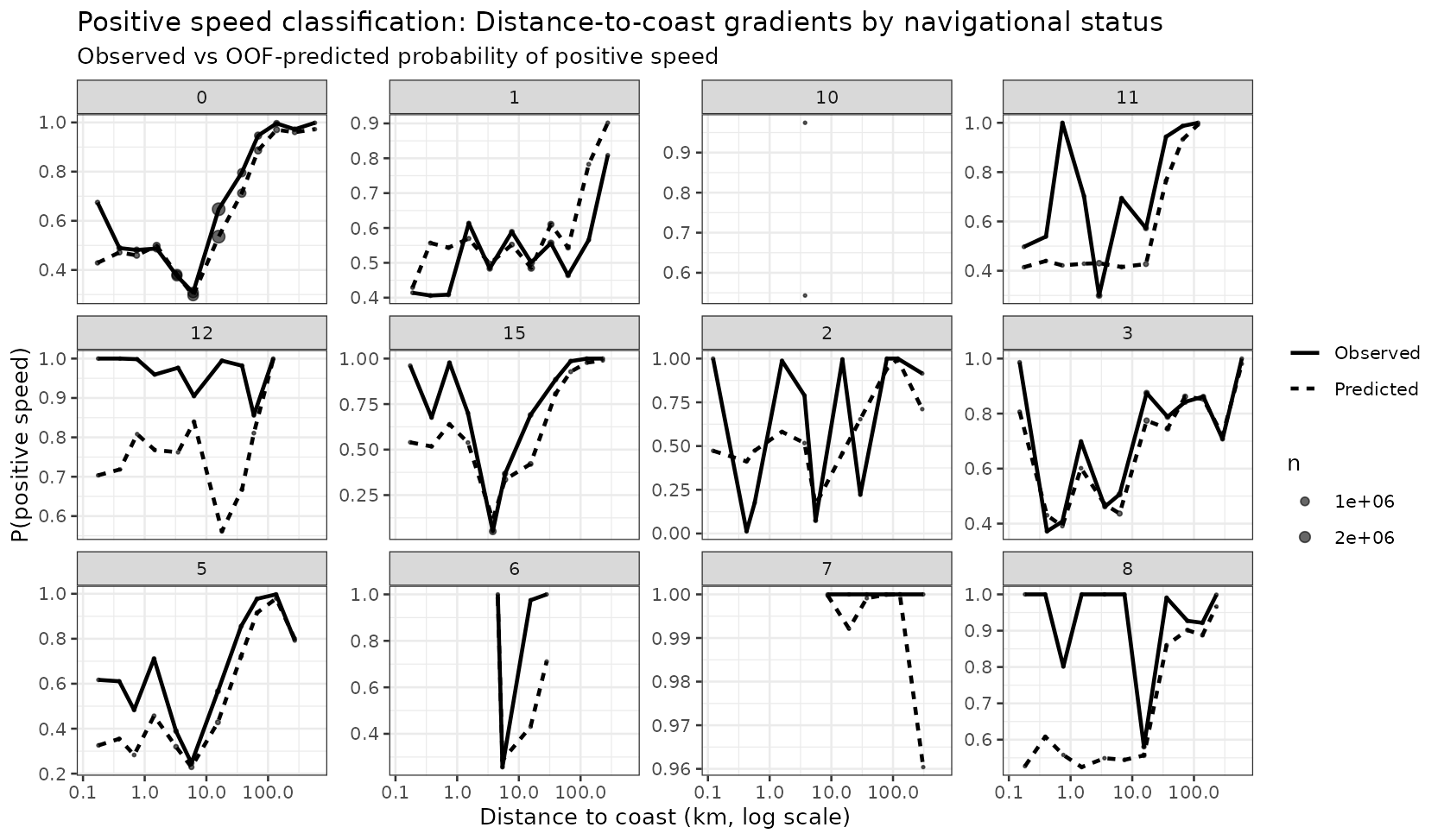}
  \caption{
  Positive SOG probability as a function of distance to coast by navigational status.
  Solid lines denote observed probabilities and dashed lines show out-of-fold predictions.
  Distance to coast is shown on a logarithmic scale.
  }
  \label{fig:clf_dist_status}
\end{figure}

Figure~\ref{fig:clf_dist_status} presents $SOG>0$ probabilities stratified by navigational
status. While baseline probabilities differ by status, most categories show a strong
increase in the likelihood of positive speed with distance offshore.
These results indicate that distance to coast interacts with navigational context rather
than acting solely as a spatial proxy.

\subsection{Positive-speed classification under wind forcing by distance to coast}

\begin{figure}[H]
  \centering
  \includegraphics[width=\textwidth]{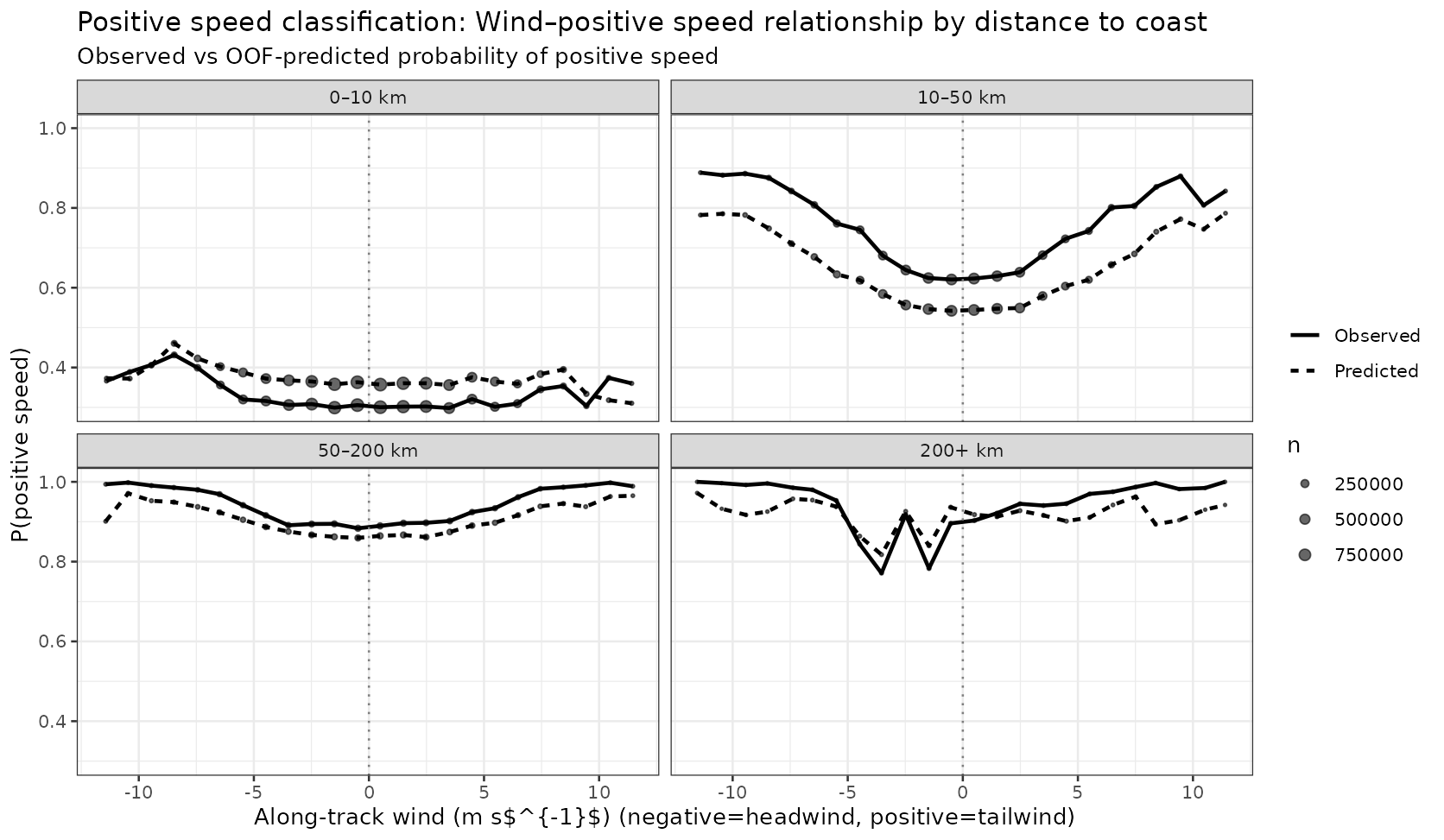}
  \caption{
  Positive $SOG>0$ probability under wind forcing by distance to coast.
  Along-track wind is defined relative to vessel heading
  (negative = headwind; positive = tailwind).
  }
  \label{fig:clf_wind_dist}
\end{figure}

Figure~\ref{fig:clf_wind_dist} shows how $SOG>0$ probability varies with along-track wind
across distance-to-coast bands. Nearshore ($<$10~km), probabilities remain low and largely
insensitive to wind, suggesting dominance of operational constraints. At intermediate
distances (10--50~km), probabilities decline under headwinds and increase under tailwinds.
Farther offshore, $SOG>0$ probabilities remain high across most wind conditions.

\subsection{Ice effects on positive speed classification by distance to coast}

\begin{figure}[H]
  \centering
  \includegraphics[width=\textwidth]{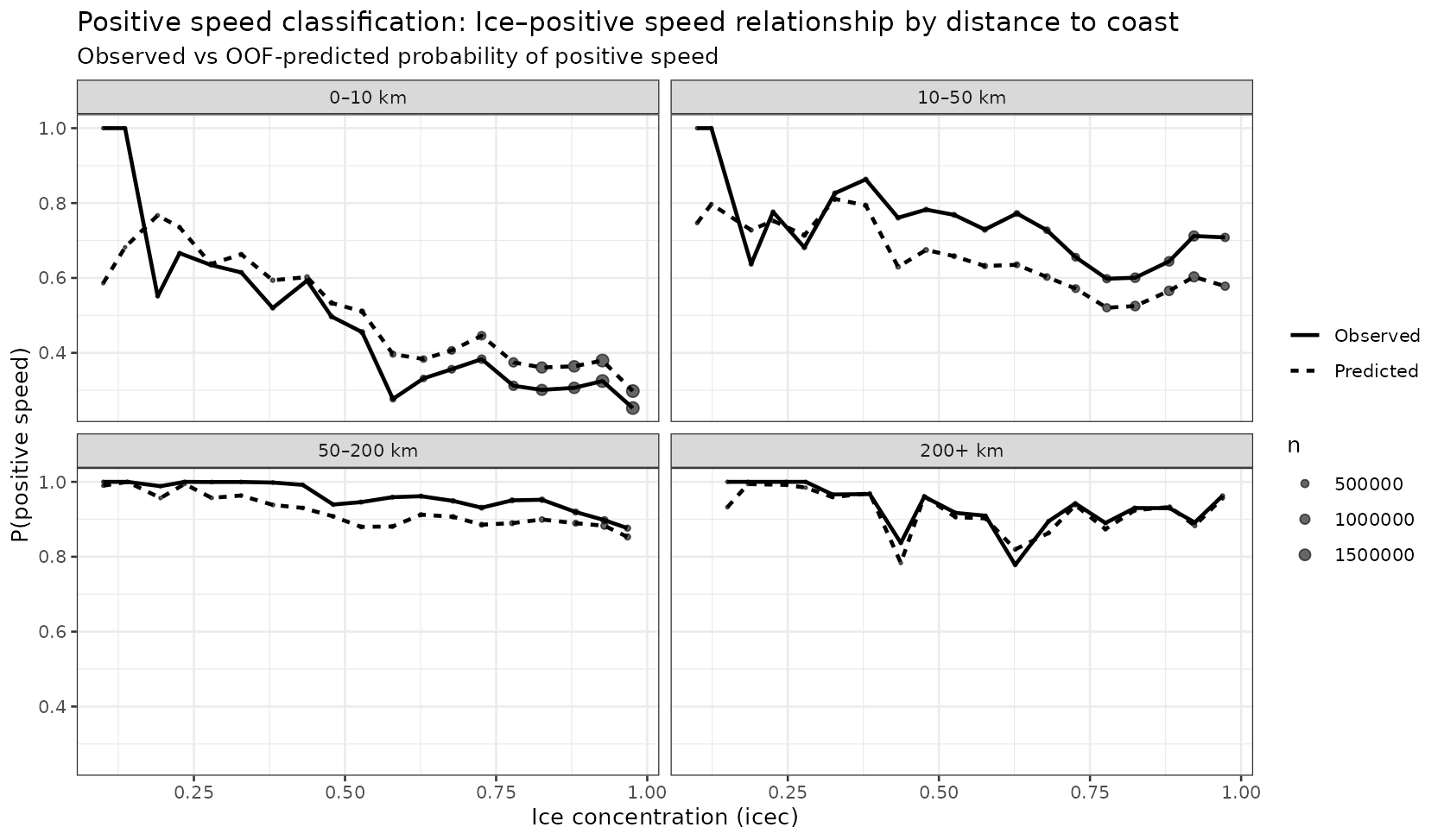}
  \caption{
  Ice $SOG>0$ probability stratified by distance to coast.
  Ice concentration is expressed as fractional coverage.
  }
  \label{fig:clf_ice_dist}
\end{figure}

Figure~\ref{fig:clf_ice_dist} shows how $SOG>0$ probability varies with sea ice
concentration across distance-to-coast bands. Nearshore, $SOG>0$ probabilities decline
sharply with increasing ice concentration. Ice effects weaken offshore, where probabilities
remain high across most ice conditions.

% ============================================================
\section{Supplementary Conditional Positive Speed Diagnostics}
\label{supp:speed}

These diagnostics examine how environmental and operational variables influence vessel
Speed Over Ground (SOG), conditional on positive movement.

\subsection{Context-dependent ice effects on vessel speed}
\label{supp:ice_diagnostics}

Although sea ice concentration exhibited modest global importance in the regression model,
stratified diagnostics indicate that ice related speed effects are systematic and strongly
context dependent.

\begin{figure}[H]
  \centering
  \includegraphics[width=\textwidth]{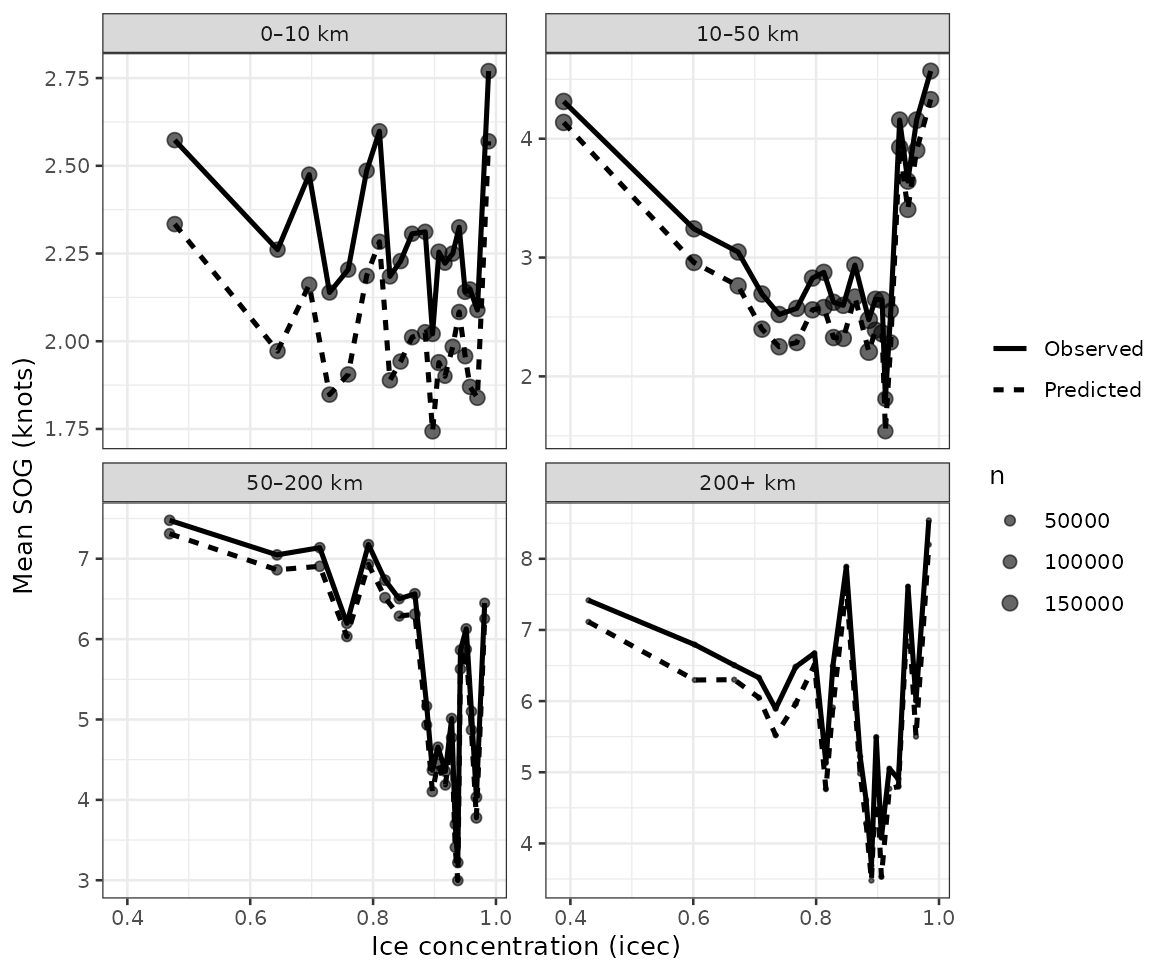}
  \caption{
  Ice-SOG relationship stratified by distance to coast.
  Solid lines show observed means and dashed lines show model predictions.
  }
  \label{fig:supp_ice_speed_stratified}
\end{figure}

Figure~\ref{fig:supp_ice_speed_stratified} presents binned mean observed and model predicted
vessel speed as a function of ice concentration across distance-to-coast bands. Speed
reductions with increasing ice concentration are most pronounced offshore (50--200~km and
200+~km), where vessels are primarily engaged in sustained transit. Nearshore regions show
lower baseline speeds and weaker ice responses, consistent with maneuvering and congestion.

At very high ice concentrations ($\mathrm{icec} \gtrsim 0.9$), increased variability reflects
sparse data support.

\subsection{Context-dependent wind effects on vessel speed}

\begin{figure}[H]
  \centering
  \includegraphics[width=\textwidth]{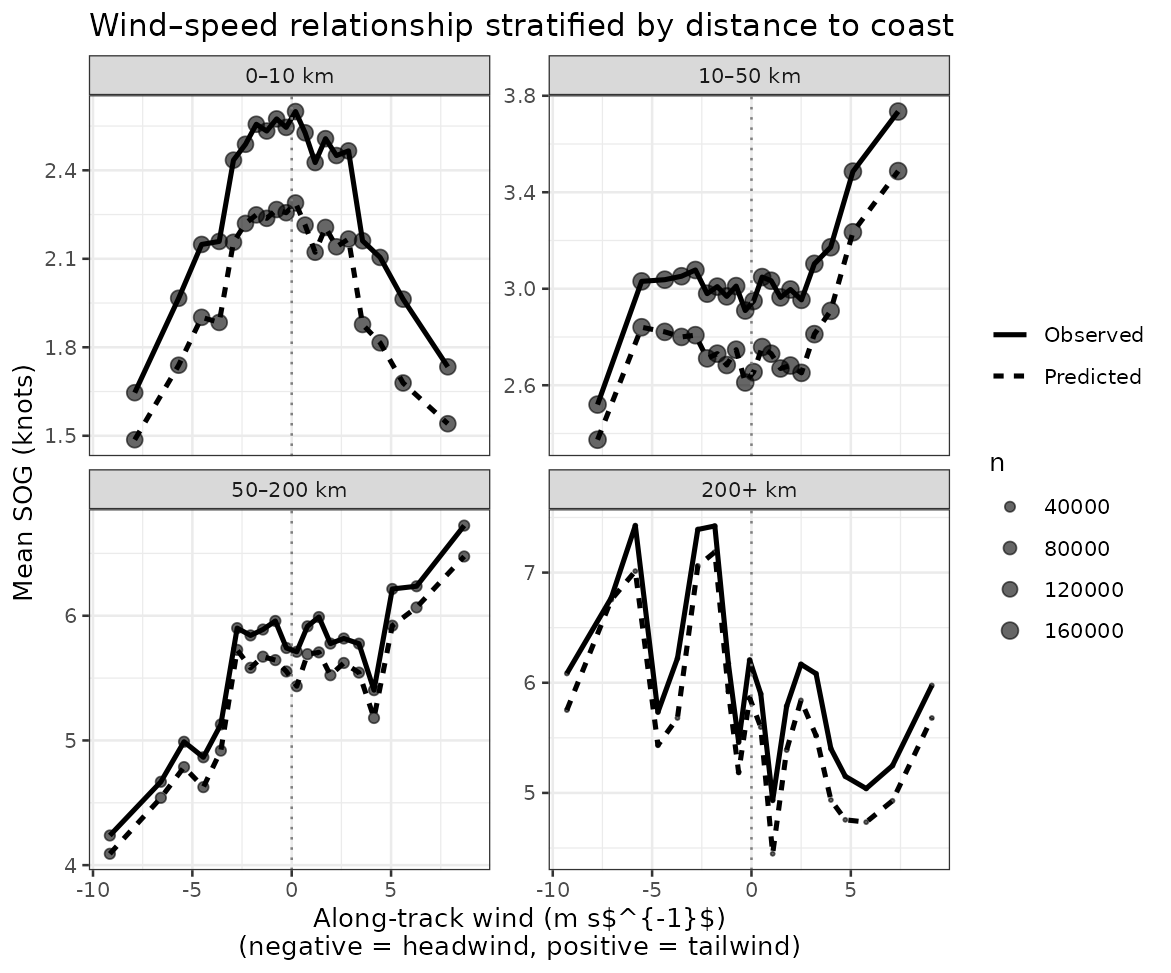}
  \caption{
  Wind--SOG relationship stratified by distance to coast.
  Negative values indicate headwinds; positive values indicate tailwinds.
  }
  \label{fig:supp_wind_speed_stratified}
\end{figure}

Figure~\ref{fig:supp_wind_speed_stratified} shows binned observed and predicted vessel speed
as a function of along-track wind. Nearshore speeds peak under near-neutral winds and
decline under both strong headwinds and tailwinds. Offshore, vessel speed increases
monotonically from headwind to tailwind conditions, consistent with sustained transit.

\subsection{Ice concentration effects on SOG by vessel group.}

\begin{figure}[H]
  \centering
  \includegraphics[width=\textwidth]{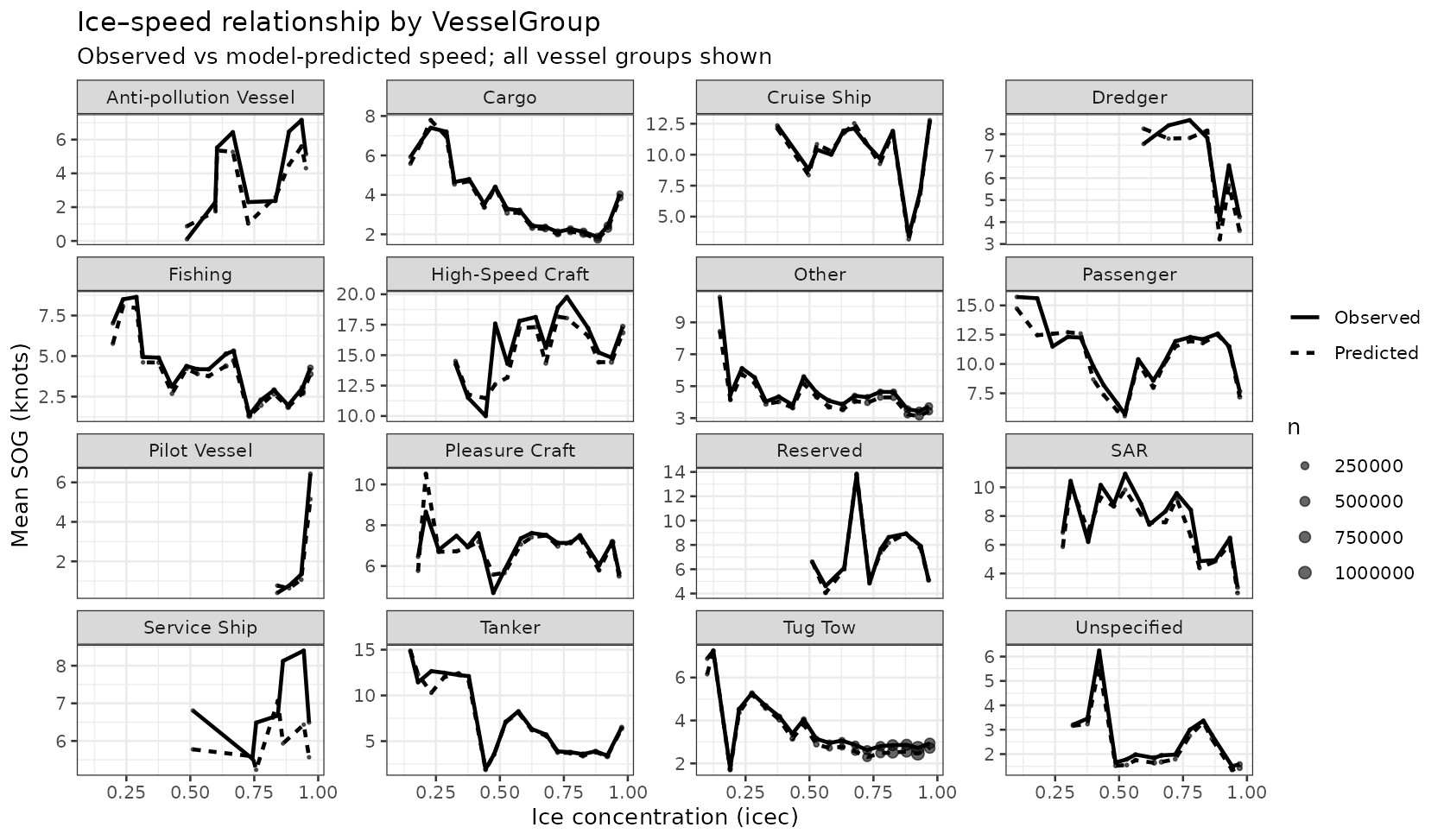}
  \caption{Ice concentration effects on SOG by vessel group.}
  \label{fig:supp_ice_speed_vesselgroup}
\end{figure}

Figure~\ref{fig:supp_ice_speed_vesselgroup} shows ice concentration effect on SOG stratified by
vessel group. Differences in slope and curvature indicate substantial heterogeneity in how
vessels adjust speed under increasing ice concentration.

\begin{figure}[H]
  \centering
  \includegraphics[width=\textwidth]{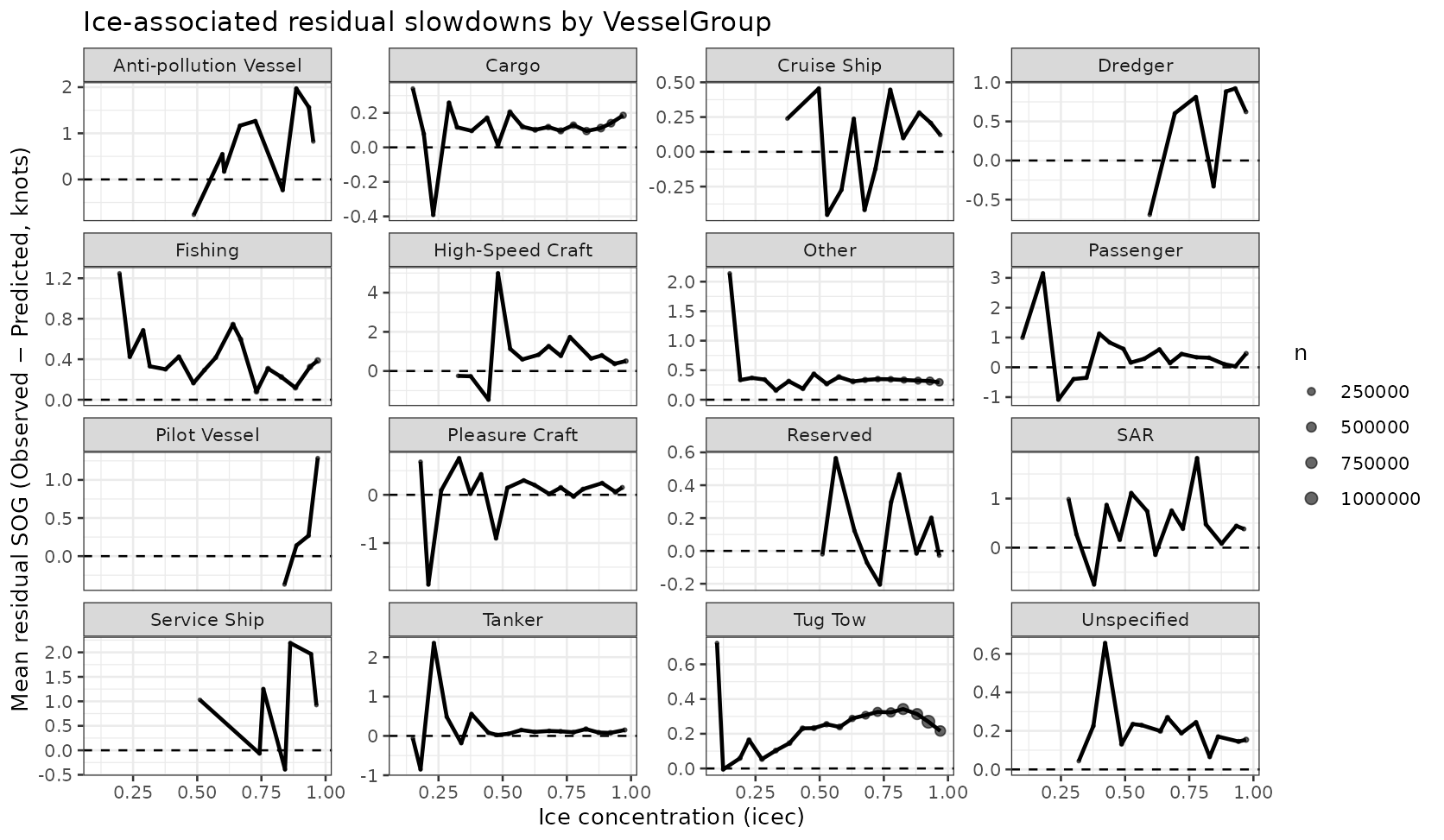}
      \caption{Distance-dependent residual speed by vessel group}
  \label{fig:supp_ice_residuals_vesselgroup}
\end{figure}

Figure~\ref{fig:supp_ice_residuals_vesselgroup} presents mean residual speeds
(observed minus predicted), highlighting vessel group specific slowdowns not fully
captured by shared covariates.

\subsection{Distance to coast effects on SOG by status.}

\begin{figure}[H]
  \centering
  \includegraphics[width=\textwidth]{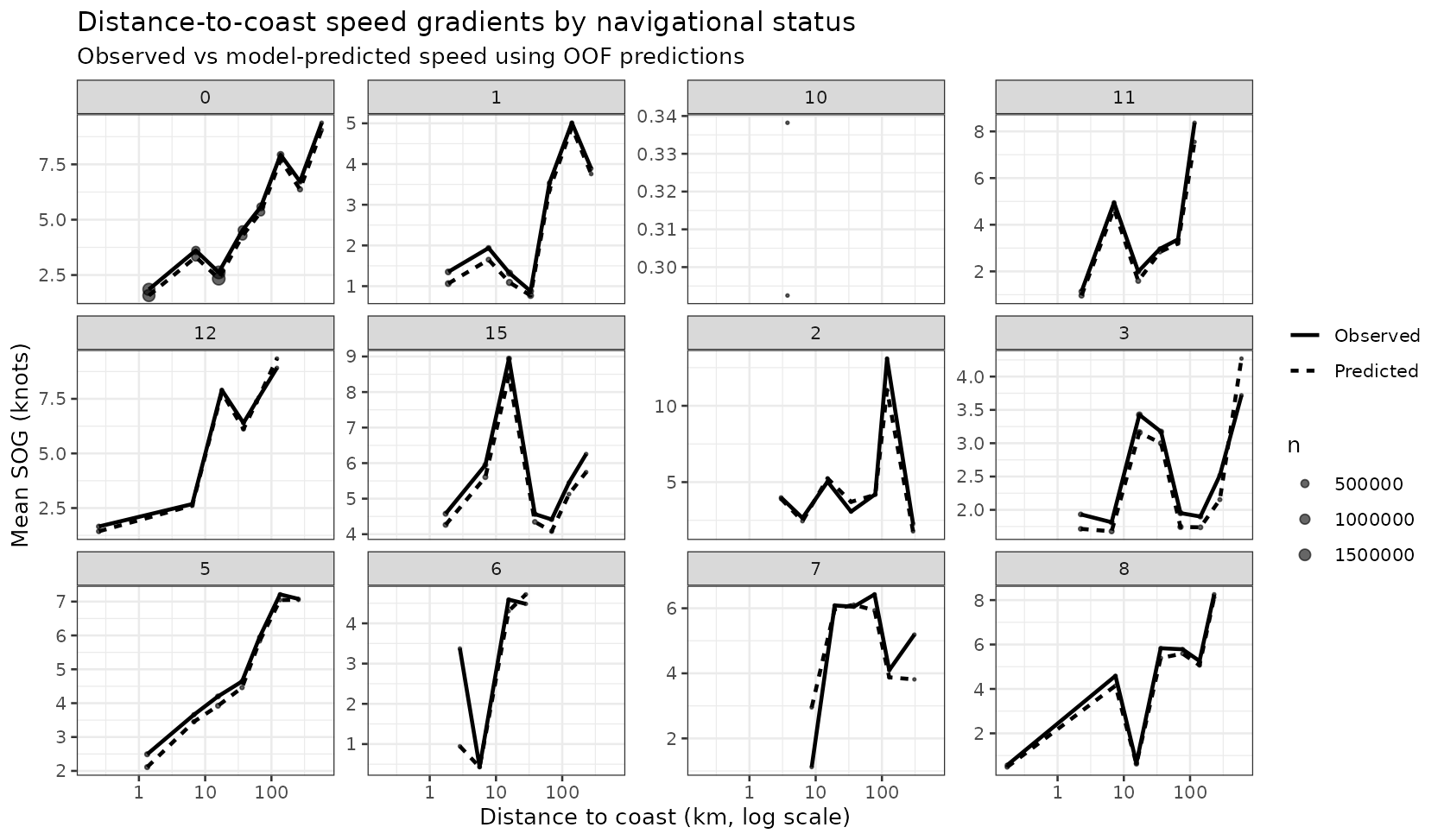}
  \caption{Distance to coast effects on SOG by status}
  \label{fig:supp_dist_speed_status}
\end{figure}

\begin{figure}[H]
  \centering
  \includegraphics[width=\textwidth]{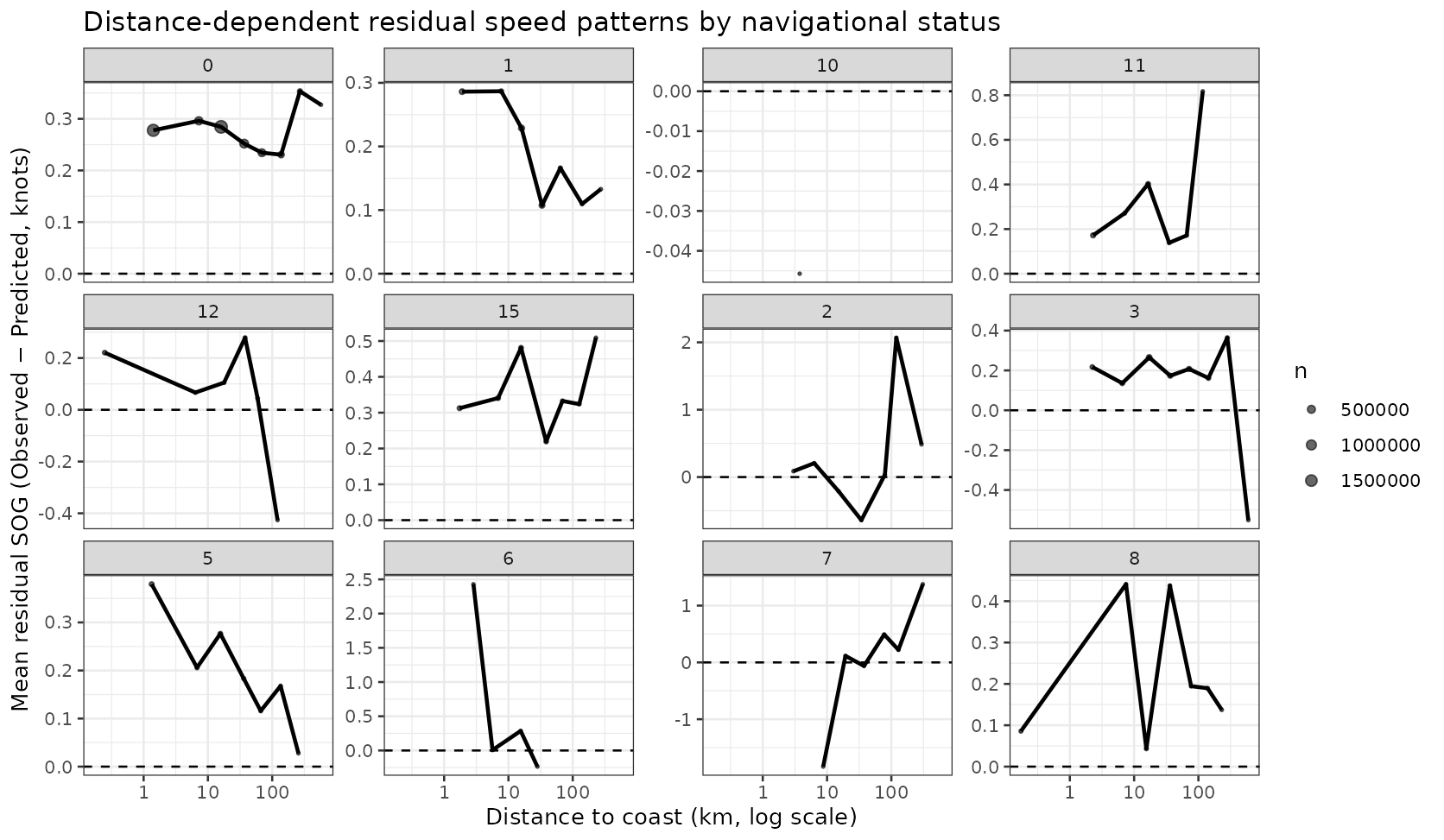}
  \caption{Distance-dependent residual speed by navigational status.}
  \label{fig:supp_dist_residuals_status}
\end{figure}

Figure~\ref{fig:supp_dist_speed_status} shows that vessel speed generally increases with
distance from shore across navigational statuses before stabilizing offshore. Residuals
(Figure~\ref{fig:supp_dist_residuals_status}) are small and centered near zero, indicating
good calibration with limited status-specific deviation.

